\documentclass[12pt]{article}

\usepackage{color} 

\usepackage{amsmath}
\usepackage{amssymb,epsfig}
\usepackage[numbers]{natbib}
\usepackage[justification=raggedright]{caption}
\usepackage{float}
\usepackage{booktabs}
\usepackage{tikz}
\usepackage{multirow}
\usepackage{subfig}
\usepackage{rotating}
\usepackage[toc,page]{appendix}
\usepackage{url}
\usepackage{tabulary}
\usepackage{ctable}
\usepackage{epsfig}
\usepackage{graphicx}
\usepackage{caption}
\usepackage{multirow}
\usepackage{amssymb}
\usepackage{subfig}

\newcommand{\bea}{\begin{eqnarray}}
\newcommand{\eea}{\end{eqnarray}}
\newcommand{\bss}{\begin{singlespace}}
\newcommand{\ess}{\end{singlespace}}		
\newcommand{\nn}{\nonumber}

\newcommand{\bfL}{{\mathbf{L}}}
\newcommand{\bfl}{{\mathbf{l}}}

\newcommand{\bfC}{{\mathbf{C}}}

\newcommand{\bftheta}{{\pmb{\theta}}}
\newcommand{\bfzeta}{{\pmb{\zeta}}}
\newcommand{\bfomega}{{\pmb{\omega}}}
\newcommand{\bflambda}{{\pmb{\lambda}}}

\newcommand{\qzero}{{{q_0}(\bfC)}}

\newcommand{\qone}{{{q_1}(\bfL^1,\bfC)}}
\newcommand{\qoneL}{{{q_1}(\bfL,\bfC)}}
\newcommand{\qonel}{{{q_1}(L^1,\bfC)}}
\newcommand{\qoneC}{{(\rho-1)\bfomega'_3\bfL^1}}

\newcommand{\qoneLC }{{(\rho-1)\bfomega'_3\bfL}}

\newcommand{\odds}{{\text{odds}}}

\newcommand{\expit}{{\text{expit}}}
\newcommand{\Bin}{{\text{Bin}}}

\newcommand{\lm}{{l(\bfL^1,\bfC;\pmb{\omega})}}
\newcommand{\lmL}{{l(\bfL,\bfC;\pmb{\omega})}}

\newcommand{\WL}{{W(S,\bfL,\bfC)}}

\newcommand{\W}{{W(S,\bfL^1,\bfC)}}

\newcommand{\Wilong}{{W(S_i,\bfL_i,\bfC_i;\bflambda, \bfomega)}}

\newcommand{\Wj}{{W_j(\bflambda, \bfomega)}}

\newcommand{\Wi}{{W_i(\bflambda, \bfomega)}}
\newcommand{\pij}{{\pi_j(\bftheta)}}

\newcommand{\pii}{{\pi_i(\bftheta)}}

\newcommand{\Tj}{{\lambda_1 + \bflambda'_2\bfC_j + \rho\bfomega'_3\bfL_j }}
\newcommand{\Nj}{{\omega_1 + \bfomega'_2\bfC_j + \bfomega'_3\bfL_j}}

\newcommand{\sumi}{{\sum_{i=1}^nR_iW_i(\bflambda, \bfomega)/\pii}}
\newcommand{\sumYi}{{\sum_{i=1}^nY_iR_iW_i(\bflambda, \bfomega)/\pii}}

\newcommand{\Wderivi}{{\begin{pmatrix}
			\frac{\partial \Wi/\pii }{\partial \bflambda} &  \frac{\partial \Wi/\pii }{\partial \bfomega} &  \frac{\partial \Wi/\pii }{\partial \bftheta}
\end{pmatrix}}}

\newcommand{\lambdaomega}{{\begin{pmatrix}
			\hat{\bflambda} - \bflambda \\\hat{\bfomega} - \bfomega \\ \hat{\bftheta} - \bftheta
\end{pmatrix}}}

\newcommand{\op}{{\mathbf{o}_p(1)}}

\newcommand{\bfU}{{\mathbf{U}}}
\newcommand{\bfS}{{\mathbf{S}}}

\usepackage{tabularx}

\usepackage{makecell}

\newcommand\independent{\protect\mathpalette{\protect\independenT}{\perp}}
\def\independenT#1#2{\mathrel{\rlap{$#1#2$}\mkern2mu{#1#2}}}

\newsavebox\smileysmall                                     \savebox\smileysmall    {                                                                 
	\begin{tikzpicture}[scale=1.2]
	         	\draw circle (2mm);
		\node[fill,circle,inner sep=0.5pt] (left eye) at (135:0.8mm) {};
		\node[fill,circle,inner sep=0.5pt] (right eye) at (45:0.8mm) {};
		\draw[thick] (-145:0.9mm) arc (-120:-60:1.5mm);                                     
	\end{tikzpicture}             
}

\newsavebox\frowniesmall                                     \savebox\frowniesmall      {                                                                 
	\begin{tikzpicture}[scale=1.2]
	\draw circle (2mm);
	\node[fill,circle,inner sep=0.5pt] (left eye) at (135:0.8mm) {};
	\node[fill,circle,inner sep=0.5pt] (right eye) at (45:0.8mm) {};
	\draw[thick] (-145:0.9mm) arc (120:60:1.5mm);                                     
	\end{tikzpicture}             
}

\usepackage{authblk} 

\title{A novel estimand  to adjust  for rescue treatment \\ in clinical trials} 
\date{}
\author[1]{Hege Michiels}
\author[2]{Cristina Sotto}
\author[2]{An Vandebosch}
\author[1,3]{Stijn Vansteelandt}
\affil[1]{Department of Applied Mathematics, Computer Science and Statistics, Ghent University, Ghent, Belgium}
\affil[2]{Janssen R\&D, a division of Janssen Pharmaceutica NV, Beerse, Belgium}
\affil[3]{Department of Medical Statistics, London School of Hygiene and Tropical Medicine, London, UK}

\begin{document}
		\maketitle

	\section*{Abstract}
	The interpretation of randomised clinical trial results is often complicated by intercurrent events. For instance, rescue medication is sometimes given to patients in response to worsening of their disease, either in addition to the randomised treatment or in its place. The use of such medication complicates the interpretation of the intention-to-treat analysis. In view of this, we propose a novel estimand defined as the intention-to-treat effect that would have been observed, had patients on the active arm been switched to rescue medication if and only if they would have been switched when randomised to control. This enables us to disentangle the treatment effect from the effect of rescue medication on a patient's outcome, while avoiding the strong extrapolations that are typically needed when inferring what the intention-to-treat effect would have been in the absence of rescue medication. 
	We develop an  inverse probability weighting method  to estimate this estimand under specific untestable assumptions, in view of which we propose a sensitivity analysis. We use the method for the analysis of a clinical trial conducted by Janssen Pharmaceuticals, in which chronically ill patients can switch to rescue medication for ethical reasons. 
	Monte Carlo simulations confirm that the proposed estimator is  unbiased in moderate sample sizes. 
	
 \noindent \emph{Keywords}: 
 Intercurrent events, Treatment switching, Causal inference, Post-treatment confounding, Mediation, Treatment discontinuation
	\section{Introduction}
		
	The ICH E9(R1) addendum (International Council for Harmonisation, 2019 \cite{ICH2019}) proposes a structured framework for clinical trial design, conduct, analysis and interpretation. It aligns the main objective of the clinical trial with the treatment effect to be estimated, called `estimand'. An estimand defines the target of estimation for a particular trial objective. The description of an estimand  thus reflects the clinical question of interest, considering  intercurrent events such as discontinuation of assigned treatment, use of an additional or alternative treatment and terminal events such as death. In this paper, we focus on the intercurrent event where patients start (additional) non-randomised medication because of an exacerbation of symptoms or insufficient therapeutic effect.   For example, in a trial of dietary intervention to reduce high blood pressure, a patient may be rescued by starting antihypertensive medication when his blood pressure fails to respond (White et al., 2001 \cite{white2001randomized}). Likewise, patients in oncology trials may sometimes start a new anticancer regimen before observing the endpoint of interest upon disease progression (Degtyarev et al., 2019 \cite{degtyarev2019estimands}).

	The use of  rescue medication complicates the interpretation of trial results, especially when its use is imbalanced between treatment arms, in which case it   typically reduces the observed treatment effect in an `intention-to-treat' or `treatment policy' analysis. Ignoring all data after switching to rescue medication is likely to deliver biased results since rescued patients or `switchers' form a highly selective group. One estimand described in the ICH E9(R1) addendum which aims to accommodate this uses the `hypothetical strategy', where a scenario is envisaged in which the intercurrent event would not occur, e.g.  in which the additional medication was not available. While this removes the effect of switching, thereby  delivering insight into the pure effect of the randomised treatments, this scenario may not be realistic when the additional medication must be available for ethical reasons. In addition, strong extrapolations  may be needed to infer what the intention-to-treat effect would have been in the absence of rescue medication. To lessen the extent of extrapolation, we propose a novel estimand, called `balanced estimand', and define it as  the intention-to-treat effect that would have been observed, had patients on the active arm been switched to rescue medication if and only if they had been switched when randomised to control. As such, we are able to distinguish the effect of the experimental treatment from the effect of rescue medication on a patient's outcome, by defining a treatment effect that would be observed if rescue treatment were balanced across both arms. 
	The proposed estimand is a variant of the hypothetical estimand discussed in the ICH E9(R1) addendum and corresponds to a so-called `natural direct effect' (Robins and Greenland, 1992 \cite{robins1992identifiability}) that can be expressed in terms of potential outcomes (Rubin, 1974 \cite{rubin1974estimating}). 
	
Existing methods for natural direct effects are not readily applicable, however, because the association between rescue treatment and outcome is typically confounded by  variables, such as disease severity, which are themselves affected by the treatment. Such post-treatment confounding poses major challenges (Daniel et al., 2015 \cite{daniel2015causal}). We accommodate this via a novel sensitivity analysis method. In particular, we propose 
 a novel inverse probability weighting approach for estimating this effect in settings where the decision to switch to rescue medication is made at one pre-specified time point. 

	  We use the proposed method for the analysis of a clinical trial conducted by Janssen Pharmaceuticals, in which chronically ill patients can switch to rescue medication for ethical reasons.  
	In addition, we perform Monte Carlo simulations to investigate the finite sample performance of the estimator.

	\section{Setting}
	\label{Section: setting}
	Suppose the data consist of independent and identically distributed observations $\{(Y_i,S_i, \\\bfL_i, R_i,\bfC_i): i = 1, \dots, n\}$, where $\mathbf{C}$ represents measured baseline covariates, $Y$ the outcome and $R$ the randomised treatment which is coded 1 for patients assigned to active treatment and 0 for those assigned to control. Variable $Y$ can represent a binary or continuous outcome, but we assume that it is not a censored time-to-event (survival) outcome.  In addition, $S$ is a binary variable indicating whether the patient switched to rescue treatment during the study ($S = 1$) or not ($S = 0$). This rescue medication can either be taken in addition to the randomised treatment or in its place. In this paper, we consider the simple setting where patients can only switch at one pre-specified time point in the trial.   We assume that whether or not the patient switches during the study is  determined by clinicians based on the information contained in the baseline covariates $\bfC$ (e.g. age) along with the possibly high-dimensional post-treatment covariates $\bfL$ (e.g. disease severity).    Within the `counterfactual' or `potential outcomes' framework (Rubin, 1974 \cite{rubin1974estimating}; Robins, 1986 \cite{robins1986new}; Pearl, 1995 \cite{Pearl1995causal}), we let $Y^r$ denote the potential outcome that would have been observed under treatment $R = r$ $ (r \in \{0,1\})$ and $Y^{r s}$ the potential outcome that would have been observed under treatment $R = r$  and switching status $S=s$ $ (s \in \{0,1\})$.  Similarly, $\mathbf{L}^r$ represents the severity of disease under treatment  $R = r$ and $S^r$ the switching status if $R = r$. 
	
	Throughout this paper, we consider  a long-term prevention trial in patients who are asymptomatic at risk for developing Alzheimer's dementia, as described by Polverejan and Dragalin (2019 \cite{polverejan2019aligning}). In this simulation case study, an experimental treatment is compared to placebo in slowing cognitive decline, measured by the 
	ADCS-PACC score (Donohue et al., 2014 \cite{donohue2014preclinical}). 
	The primary endpoint is the change in  ADCS-PACC score from baseline to month 54. One of the intercurrent events  considered by the authors is the initiation of treatment with cholinesterase inhibitors during the course of the trial. This rescue medication is given if medically indicated and may deliver therapeutic benefit due to the symptomatic effects of cholinesterase inhibitors. In this example, we assume for illustrative purposes that   the change in  ADCS-PACC score from baseline to month 26 is an indicator for the severity of disease, and is used to determine whether the patient should initiate treatment with cholinesterase inhibitors. 
		
	\subsection{Handling intercurrent events}
	\label{section: handling intercurrent events}
	
	Descriptions of three different strategies to handle intercurrent events are listed below, each reflecting a different clinical question of interest in the above setting (International Council for Harmonisation, 2019 \cite{ICH2019}). 	First, in the `treatment policy strategy' or `intention-to-treat effect' $E\left(Y^{1}-Y^{0} \right)$,  the occurrence of the intercurrent event is considered irrelevant in defining the treatment effect of interest. 
	By ignoring whether or not the patient took rescue medication, the treatment effect targeted by this estimand is a combined effect of the initial randomised treatment and the treatment modified as a result   of the rescue medication.  Typically, rescue treatment leads to a reduced outcome difference between treatment regimens, which may make  it seem easier to demonstrate non-inferiority or equivalence between arms (Ratitch et al., 2020 \cite{ratitch2020choosing}).  In such cases, it may be difficult to make decisions about the efficacy of treatment regimens.
	 This is illustrated by a toy example with binary outcome in table \ref{table: toy example estimands}. It shows the potential outcomes under treatment and control  of five patients who have access to rescue medication during the trial. Some of these patients would never switch to rescue treatment, regardless of the assigned treatment,  some would always switch and some would switch only if assigned to control. Although the experimental treatment has a beneficial inherent effect on the outcome of patients, this is  not captured here by the treatment policy estimand. In particular, we illustrate that this beneficial effect can be diluted if more patients would  switch to rescue medication under control than under treatment since rescue medication positively influences the outcome. 
	
	Next, we take a closer look at the `hypothetical strategy', where the intercurrent event is seen as a mediating factor for inference about the treatment regimen of interest. The goal is to estimate the treatment effect under a hypothetical scenario, where the intercurrent event is removed, for example, what would have happen-ed if rescue medication was not available. Using the  potential outcomes framework this can be expressed as $E\left(Y^{10}-Y^{00} \right)$,
	with $Y^{r0}$ the outcome of a patient that would have been observed if (s)he was assigned to treatment $r \in \{0,1\}$ and did not switch to rescue treatment during the trial ($S = 0$). This estimand is again illustrated in table \ref{table: toy example estimands}. 
	This hypothetical strategy clearly distinguishes the effect of the randomised treatment from the effect of  rescue treatment, 
	but is prone to extrapolation when it is difficult to consider not giving rescue medication to some patients, as is the case in the Alzheimer's disease example. 
	In addition, the scenario  `if patients had no access to rescue medication' is sometimes criticized because it is incompatible with ethical conduct since patients cannot be forced to adhere (Mallinckrodt et al., 2017 \cite{mallinckrodt2017choosing}; Permutt, 2016 \cite{permutt2016taxonomy}), thereby also raising questions about the relevance of this effect  for policy making.  
	
	Finally, we consider the `principal stratification strategy', where the objective is to estimate the treatment effect 
	for the  patients who would never need to switch to rescue therapy, regardless of the assigned treatment: $E\left(Y^{1}-Y^{0} \right | S^0 = S^1 = 0)$. 
	Since for each patient  we only observe what happens on the treatment to which (s)he was randomised, it is not possible to observe which principal stratum the patient belongs to. Consequently, the principal stratification estimand represents the treatment effect for an unknown and non-identifiable group of patients. 
	In addition, the subgroup of patients that would never need rescue therapy, referred to as the `always-compliers' or `always-takers' (e.g. VanderWeele, 2011 \cite{vanderweele2011principal}), is typically a selective subgroup of the population, making it difficult to generalise conclusions to the broader level of the patient population. In our toy example in table \ref{table: toy example estimands}, the principal stratum of patients who would never switch is a small subgroup of the considered population.  Estimation of the principal stratification effect relies, like the other discussed estimands, on unverifiable assumptions. However, the assumptions necessary to identify the principal stratification effect are generally stronger since they relate to the joint distribution of the counterfactuals $S^0$ and $S^1$.

\begin{table}[h]
	\centering
	\begin{tabular}{ll m{0.5cm}m{0.5cm}m{0.5cm}m{0.5cm}m{0.5cm}l}
		\toprule
		Estimand  & \multicolumn{6}{l}{Potential outcomes}&Treatment effect estimate\\ 
		\midrule
		\multirow{2}{*}{\makecell[l]{Treatment \\ policy}} & $Y^1$ &\includegraphics[width=0.04\textwidth]{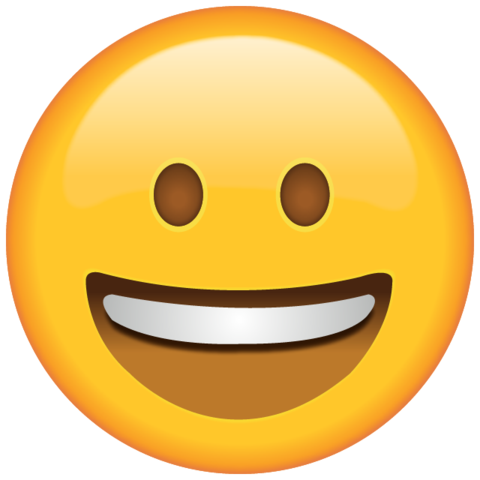} & \includegraphics[width=0.04\textwidth]{Happy.png} &\includegraphics[width=0.04\textwidth]{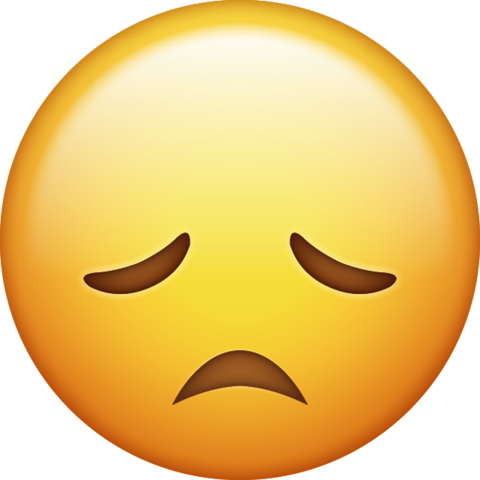} & \includegraphics[width=0.04\textwidth]{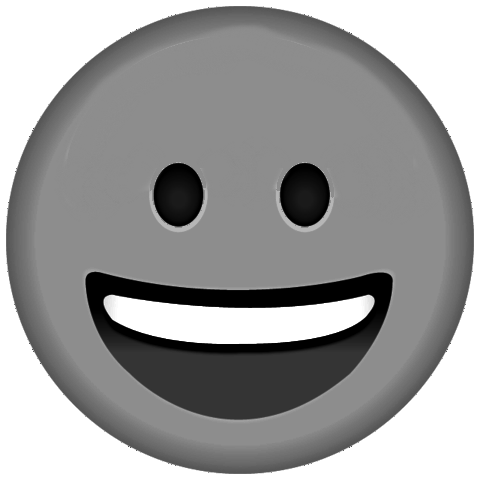} & \includegraphics[width=0.04\textwidth]{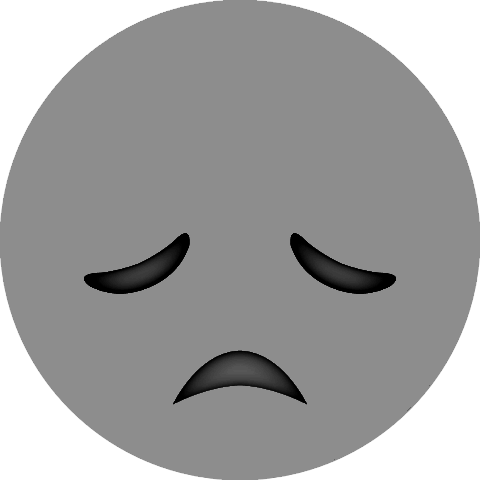}  & \multirow{2}{*}{\makecell[l]{$E(Y^1-Y^0)$ \\ $ = 3/5-3/5$\\ $=0$}} \\[8pt]  
		& $Y^0$ & \includegraphics[width=0.04\textwidth]{Happy.png} & \includegraphics[width=0.04\textwidth]{Sad.png} &\includegraphics[width=0.04\textwidth]{HappyGrey.png} & \includegraphics[width=0.04\textwidth]{HappyGrey.png} & \includegraphics[width=0.04\textwidth]{SadGrey.png}  \\
		\cmidrule(ll){1-8} 	
		Hypothetical & $Y^{10}$ & \includegraphics[width=0.04\textwidth]{Happy.png} & \includegraphics[width=0.04\textwidth]{Happy.png} &\includegraphics[width=0.04\textwidth]{Sad.png} & \includegraphics[width=0.04\textwidth]{Happy.png} & \includegraphics[width=0.04\textwidth]{Sad.png} &   \multirow{2}{*}{ \makecell[l]{$E(Y^{10}-Y^{00})$ \\ $ =3/5-2/5$ \\ $=1/5 $}}\\[8pt]  
		& $Y^{00}$ & \includegraphics[width=0.04\textwidth]{Happy.png} & \includegraphics[width=0.04\textwidth]{Sad.png} &\includegraphics[width=0.04\textwidth]{Sad.png} & \includegraphics[width=0.04\textwidth]{Happy.png} & \includegraphics[width=0.04\textwidth]{Sad.png}  \\
		\cmidrule(ll){1-8} 
		\multirow{2}{*}{\makecell[l]{Principal \\ stratification}} & $Y^{1}$ & \includegraphics[width=0.04\textwidth]{Happy.png} & \includegraphics[width=0.04\textwidth]{Happy.png} & &  & & \multirow{2}{*}{ \makecell[l]{ $E(Y^{1}-Y^{0}  | S^0 = S^1 = 0)$\\ $= 2/2-1/2$ \\$=1/2 $ }}\\[8pt]  
		& $Y^{0}$ & \includegraphics[width=0.04\textwidth]{Happy.png} & \includegraphics[width=0.04\textwidth]{Sad.png} & &  &   \\
		\cmidrule(ll){1-8} 
		Balanced& $Y^{1S^0}$ & \includegraphics[width=0.04\textwidth]{Happy.png} & \includegraphics[width=0.04\textwidth]{Happy.png} &\includegraphics[width=0.04\textwidth]{HappyGrey.png} & \includegraphics[width=0.04\textwidth]{HappyGrey.png} & \includegraphics[width=0.04\textwidth]{SadGrey.png} & \multirow{2}{*}{\makecell[l]{$E(Y^{1S^0}-Y^0)$ \\ $ =4/5-3/5$ \\ $=1/5 $}} \\[8pt]  
		& $Y^0$ & \includegraphics[width=0.04\textwidth]{Happy.png} & \includegraphics[width=0.04\textwidth]{Sad.png} &\includegraphics[width=0.04\textwidth]{HappyGrey.png} & \includegraphics[width=0.04\textwidth]{HappyGrey.png} & \includegraphics[width=0.04\textwidth]{SadGrey.png}  \\				
		\bottomrule
	\end{tabular}
	\caption{Toy example with five patients to illustrate  the  estimands discussed using the potential outcomes. The outcome $Y$ is binary. Meaning of the smileys: \\
	\protect\includegraphics[height=0.5cm]{Happy.png}:  $Y = 1$ (success) and patient did not switch to rescue \\
	\protect\includegraphics[height=0.5cm]{Sad.png}:  $Y = 0$ (failure) and patient did not switch to rescue \\
	\protect\includegraphics[height=0.5cm]{HappyGrey.png}:  $Y = 1$ (success) and patient did  switch to rescue \\
	\protect\includegraphics[height=0.5cm]{SadGrey.png}:  $Y = 0$ (failure) and patient did  switch to rescue 
	}
	\label{table: toy example estimands}
\end{table}

	\subsection{Balanced estimand}
	\label{Section: Natural direct effect}
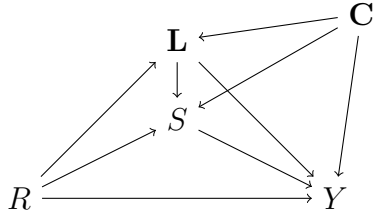
\begin{figure}[h]
		\centering
		\begin{tikzpicture}[scale=0.7]
		\path (0,0) node(r) {$R$}
		(6,0) node(y) {$Y$}
		(3,3) node(l) {$\mathbf{L}$}
		(6.5,3.5) node(c) {$\mathbf{C}$}
		(3,1.5) node(s) {$S$};
		\draw[->,black] (r) --(1.5,0.75)node[above] {}-- (s);
		\draw[->,black] (r) --(1.5,1.5)node[above] {}-- (l);
		\draw[->,black] (r) --(4,0)node[above] {}-- (y);
		\draw[->,black] (s) --(4.5,0.75)node[above] {}-- (y);
		\draw[->,black] (l) --(3,2.25)node[right] {}-- (s);
		\draw[->,black] (l) --(4.5,1.5)node[above] {}-- (y);
		\draw[->,black] (c) --(4.75,3.25)node[above] {}-- (l);
		\draw[->,black] (c) --(6.25,1.75)node[right] {}-- (y);
		\draw[->,black] (c) --(4.75,2.5)node[above] {}-- (s);
		\end{tikzpicture}
		\caption{Causal diagram with treatment $R$, outcome $Y$, severity of disease $\bfL$, switching status $S$ and baseline covariates $\bfC$.  }
		\label{fig: causal diagram}
	\end{figure}

	To lessen the extent of extrapolation, we here propose a novel estimand, defined as the treatment effect that would have been observed in the setting where the use of rescue treatment were balanced across  both arms. 
		In particular, 
	we consider the intention-to-treat effect that would have been observed, had patients on the active arm been switched to rescue medication if and only if they would have been switched when randomised to control.   By doing this, we disentangle the effect of rescue medication  from the effect of the randomised treatment. 
	This is tantamount  to disentangling  the treatment pathways  $R \rightarrow Y$ and $R\rightarrow \bfL \rightarrow Y$  from the switching pathways $R\rightarrow S \rightarrow Y$ and $R\rightarrow \bfL \rightarrow  S \rightarrow Y$   in the causal diagram in figure \ref{fig: causal diagram}. 	
 
	Within the potential outcomes framework, we let $Y^{rS^0}$ denote the potential outcome that would have been observed for a patient if (s)he was assigned to treatment $r \in \{0,1\}$, but would switch as under control.  For a  patient who would need to switch to rescue treatment if s(he) was assigned to control, $S^0$ equals 1 and consequently, $Y^{1S^{0}} = Y^{11}$ represents the outcome that would have been observed if this patient was assigned to treatment, but switched to rescue medication during the trial. 
	Similarly, for a patient who would not  switch to rescue treatment if s(he) was assigned to control, $S^0$ equals 0 and $Y^{1S^{0}} = Y^{10}$ represents the outcome that would have been observed if this patient was assigned to treatment and stayed on treatment during the entire trial. 
	Finally, $Y^{0S^{0}}$ is the outcome of a patient if (s)he was in the control group and would switch as under control. Therefore, $Y^{0S^{0}}$ can also be written as $Y^0$ 
	, the outcome that would have been observed if the patient was assigned to the control arm of the trial. 
	Using this notation, the proposed treatment effect  can be expressed as the natural direct effect (Robins and Greenland, 1992 \cite{robins1992identifiability})  
	\begin{align}
	\mu := E\left(Y^{1S^{0}}-Y^{0S^{0}}\right) 
	= E\left(Y^{1S^{0}}-Y^{0} \right). 
	\label{IE placebo}
	\end{align}
	The toy example in table \ref{table: toy example estimands} illustrates that this balanced estimand defines an effect for the entire study population, with the use of rescue treatment being balanced across both arms. In the context of the Alzheimer's dementia study, 
	 $E(Y^{1S^{0}})$ is the average change in cognitive score if all patients would be assigned to experimental treatment but would initiate rescue medication only if they would do this under placebo. By doing so,  we envisage a realistic hypothetical scenario where patients have access to rescue medication, but its mediating effect is removed by keeping the patients who switch fixed across both arms. It can be defined according to the guidelines of the  ICH E9(R1) addendum (International Council for Harmonisation, 2019 \cite{ICH2019}), as shown in appendix \ref{Appendix: estimands framework}.

	The proposed balanced estimand (\ref{IE placebo}) corresponds to the natural direct effect where switching is fixed to the natural value it would have been under control. However, in some situations (see later), it may be more interesting or relevant to fix the switching status to the natural value it would have been under experimental treatment: $E\left(Y^{1}-Y^{0S^1} \right)$. This natural direct effect  can  be estimated using the methods described for estimating (\ref{IE placebo}) below, upon interchanging the meaning of $R=0$ and $R=1$.

		\section{Identification}
		\label{Section: Identification}
		
		For simplicity, we  assume  randomisation to be independent of  baseline covariates, i.e. $R \independent \bfC$. However, if randomisation depends on certain baseline covariates, as is the case under stratified randomisation, the proposed method can easily be adjusted by replacing the marginal probability $P(R)$ by the conditional probability $P(R|\bfC)$ in the identification results below (see appendix \ref{Appendix: identification}).
		Identification of $\mu$ relies on consistency and randomised assignment which implies 
		$R \independent Y^{r} $ and $R \independent (Y^{1s}, S^0, \bfL^1, \bfC)$ for $ r,s \in \{0,1\}$. 
		The second part of   effect (\ref{IE placebo}), $\mu_0 :=E(Y^0)$, can then simply be identified as the average outcome in the control arm: 
	\begin{align*}
	\mu_0 =E(Y^{0}) 
	=E(Y^{0}|R = 0) 
	= E(Y| R = 0).
	\end{align*}

	The first part of  effect (\ref{IE placebo}),  $\mu_1 := E\left(Y^{1S^{0}}\right)$, 
	cannot be identified from the observed data without making untestable assumptions since the counterfactual outcomes $Y^{1S^{0}}$ are not observable. 
	In particular, we will assume that the potential decision to switch $S^r$ can be fully attributed to the baseline covariates and severity of disease  in the sense that it has no residual dependence on the potential outcomes $Y^{1s}$ i.e., $S^r \independent Y^{1s}  | \bfL,  \bfC$ ($s,r \in \{0,1\}$)
	 (see appendix \ref{Appendix: identification} for further details).
	 Under these assumptions,  $\mu_1$ can be rewritten as a weighted average outcome among the patients in the treatment group:
	\begin{align}
	\mu_1 &= E\left(Y^{1S^{0}}\right) \nonumber\\
	&= \int E(Y^{1 s}|\bfL^1 = \bfl,\bfC) P(S^0 = s|\bfL^1 = \bfl, \bfC) f(\bfL^1 = \bfl|\bfC) f(\bfC) ds d\bfl d\bfC \nonumber \\
	&= \int E(Y|S=s,\bfL = \bfl, R=1,\bfC) P(S=s|\bfL^1 = \bfl,R=0,\bfC)f(\bfL = \bfl|R=1,\bfC)f(\bfC) ds d\bfl d\bfC \nonumber \\
	&= E\bigg[Y\frac{R}{P(R=1)}\W\bigg],  \label{eq: definition mu1}
	\end{align}
	with weights
	\begin{align*}
	\W &:= \frac{P(S|\bfL^1,R=0,\mathbf{C})}{P(S|\bfL^1,R=1,\mathbf{C})}. 
	\end{align*}
	Identity (\ref{eq: definition mu1}) shows that the above assumptions are not sufficient to identify $\mu_1$ because $P(S|\bfL^1,R=0,\bfC)$ is not identified. This can also be understood upon rewriting $E\left(Y^{1S^{0}}\right)$ as $E\left(Y^{1\bfL^1S^{0\bfL^0}}\right)$, which shows that identification requires information on the joint distribution of $\bfL^0$ and $\bfL^1$, given $\bfC$ (Daniel et al., 2015 \cite{daniel2015causal}). It is precisely this which has hindered the development of natural direct effect estimates in the presence of post-treatment confounders $\bfL$ (VanderWeele et al., 2014 \cite{vanderweele2014effect}). 
	Existing solutions have either added the pathway $R \rightarrow \bfL \rightarrow S \rightarrow Y$ to the direct treatment effect $R \rightarrow Y$, considered a hypothetical scenario where the decision to switch is made independently of $\bfL$ (Vansteelandt and Daniel, 2017 \cite{vansteelandt2017interventional}) or considered a setting where $\bfL^0$ is independent of $\bfL^1$, given $\bfC$ (Robins and Richardson, 2010 \cite{robins2010alternative}).
	 All of these are undesirable in our setting. In particular, we argue that a large part of the treatment effect will arise by $R$ having an effect on $\bfL$. Therefore, we aim to distinguish  the direct pathways $R \rightarrow Y$ and $R\rightarrow \bfL \rightarrow Y$  from the others. In view of this, we develop here  a novel approach. 
	
	 Given that $\bfL$ is potentially high-dimensional, we will herein avoid modelling the  joint distribution of $\bfL^0$ and $\bfL^1$. We will instead make assumptions that are sufficient to identify the probability to switch a control patient, conditional on $\bfL^1$, i.e.  $P(S=1|\bfL^1,R=0,\bfC)$. In particular, since the decision to switch a control patient is based on his/her observed health status  $\bfL^0$, which will often be strongly correlated with $\bfL^1$, we will introduce a `dilution factor' $\rho$ which expresses to what extent the association between switching $S$ and severity of disease under treatment $\bfL^1$ is weaker in the control group than in the treatment group. 
Specifically, assuming that	 $P(S=1|\bfL^1,R=1,\bfC) = \text{expit}(\omega_1 + \bfomega_2'\bfC + \bfomega_3'\bfL^1)$, we  
will model $P(S = 1|\bfL^1,R=0,\mathbf{C})$ as
\begin{equation}
	P(S = 1|\bfL^1,R=0,\mathbf{C}) = \expit(\lambda_1 + \bflambda'_2\bfC + \rho\bfomega'_3\bfL^1),
	\label{eq: model q link prob S}
\end{equation} 
 for given $\rho$,
with unknown parameter values $\lambda_1$ and $\bflambda'_2$. In appendix \ref{Appendix: sensitivity analysis}, we show a data generating mechanism under which this model is correctly specified.
In appendix \ref{Appendix: identification}, we propose a   more general model and discuss nonparametric identification.

The sum $\lambda_1 - \omega_1 + (\bflambda_2 - \bfomega_2)' \bfC + (\rho-1)\bfomega_3'\bfL^1$ can be interpreted as the log odds ratio for switching in the control group versus the treatment group: 
\begin{align*}
\exp\{\lambda_1 - \omega_1 + (\bflambda_2 - \bfomega_2)' \bfC + (\rho-1)\bfomega_3'\bfL^1\} &= \frac{\odds(S=1|R=0, \bfL^1,\bfC)}{\odds(S=1|R=1, \bfL^1,\bfC)}. 
\end{align*}
Here, $\qoneC$ represents the extent to which the association between switching and severity of disease under treatments differs between the two treatment groups: 
\begin{align}
\exp\{\qoneC\} &= \frac{\odds(S=1|R=0, \bfL^1,\bfC)/\odds(S=1|R=1, \bfL^1,\bfC)}{\odds(S=1|R=0, \bfL^1=\mathbf{0},\bfC)/\odds(S=1|R=1, \bfL^1=\mathbf{0},\bfC)}.
\label{eq: meaning q1}
\end{align}
Here, $\bfL^1 = \mathbf{0}$ is a reference value, e.g. not severely ill under treatment. The parameter $\rho$ needs to be specified by the user and  can be used as a sensitivity parameter by repeating the estimation of $\mu_1$ for a range of values $\rho$.  Since $\rho$ equals the correlation between $\bfL^0$ and  $\bfL^1$ conditional on $\bfC$ under certain data generating mechanisms (see appendix \ref{Appendix: sensitivity analysis}),  varying $\rho$ over the interval $[0.8,1]$ will often constitute a good choice.  In particular, $\rho = 1$ implies 
\begin{equation*}
\frac{\odds(S=1|R=0, \bfL^1,\bfC)}{\odds(S=1|R=1, \bfL^1,\bfC)} = \frac{\odds(S=1|R=0, \bfL^1=\mathbf{0},\bfC)}{\odds(S=1|R=1, \bfL^1=\mathbf{0},\bfC)},
\end{equation*}
in which case differences between the switching statuses in both treatment groups can be fully attributed to the baseline covariates. This is generally implausible, suggesting that $\rho = 1$ can be viewed as an upper bound.

	\section{Inverse probability weighting  (IPW) estimator}
	\label{Section: estimation}

	\label{Section: IPW} 
	
	From (\ref{eq: definition mu1}), it follows that $\mu_1$ can be estimated by calculating a weighted average of the outcome of the treated patients using weights 
	$\frac{P(S|\bfL^1,R=0,\bfC)}{P(S|\bfL^1,R=1,\bfC)}$.
	This motivates the following approach:
	\begin{enumerate}
		\item Fit a parametric model 
		for the probability of switching in the treatment group:
		$P(S = 1|\bfL,R = 1,\bfC) = \text{expit}(\omega_1 + \bfomega_2'\bfC + \bfomega_3'\bfL)$.
		\item  
		Estimate parameter $\bflambda = (\lambda_1, \bflambda_2)'$ in model $P(S = 1|\bfL^1,R=0,\mathbf{C}) = \expit(\lambda_1 + \bflambda'_2\bfC + \rho\bfomega'_3\bfL^1)$ by solving the following estimating equations: 
	\begin{eqnarray*}
		&&\mathbf{0}	= \sum_{i = 1}^n \begin{pmatrix}
			1 \\ \bfC_i
		\end{pmatrix}  \left( \frac{(1-R_i)(1-S_i)}{1-\hat{\pi}} \right. \\
		&& \left. -\frac{1}{\hat{\pi}}  \frac{R_i(1-S_i)}
		{\text{expit}(\hat{\omega}_1 + \hat{\bfomega}_2'\bfC_i + \hat{\bfomega}_3'\bfL_i)\left\{\exp\left(\lambda_1 - \hat{\omega}_1+ (\bflambda_2 -\hat{\bfomega}_2)'\bfC_i  + (\rho-1)\hat{\bfomega}_3'\bfL_i \right)-1 \right\} +1 }\right) ,
	\end{eqnarray*} 
		with $\hat{\pi} = n^{-1}\sum_{i = 1}^n R_i$ the sample mean of $R$ and $\hat{\omega}_1$, $\hat{\bfomega}_2$ and $\hat{\bfomega}_3$  the estimators for ${\omega}_1$, ${\bfomega}_2$ and ${\bfomega}_3$ obtained in step  1. 
		\item Estimate the weights $\WL$: 
		\begin{equation*}
		W(1,\bfL,\bfC;\hat{\bflambda}, \hat{\bfomega} ) = 	\frac{\expit(\hat{\lambda}_1+\hat{\bflambda}_2'\bfC+\rho\hat{\bfomega}_3'\bfL)}{\expit(\hat{\omega}_1 + \hat{\bfomega}'_2\bfC+ \hat{\bfomega}'_3\bfL )}
		\end{equation*}	
		for  switchers, and 
		\begin{equation*}
		W(0,\bfL,\bfC;\hat{\bflambda}, \hat{\bfomega} ) = 	\frac{1-\expit(\hat{\lambda}_1+\hat{\bflambda}_2'\bfC+\rho\hat{\bfomega}_3'\bfL)}{1-\expit(\hat{\omega}_1 + \hat{\bfomega}'_2\bfC+ \hat{\bfomega}'_3\bfL )} 
		\end{equation*}	
		for  non-switchers, 
		with 	$\hat{\lambda}_1$ and $\hat{\bflambda}_2$ the estimates for ${\lambda}_1$ and ${\bflambda}_2$ obtained in step 2.
		\item Estimate $\mu_1$ as the weighted average outcome  
    	\begin{equation*}
		\hat{\mu}_1 = \frac{ \sum_{i = 1}^n 	Y_i{R_i}W_i}{\sum_{i = 1}^n 	{R_i}W_i}, 
		\end{equation*}
		with $W_i = W(S_i,\bfL_i,\bfC_i;\hat{\bflambda}, \hat{\bfomega} )$.
		
	  	\item Estimate $\mu_0$ as the weighted  average outcome
	  	\begin{equation*}
	  	\hat{\mu}_0 = \frac{\sum_{i = 1}^n Y_i (1-R_i)}{\sum_{i = 1}^n 1-R_i}.
	  	\end{equation*}
	\end{enumerate} 
	Finally, $\mu$ is estimated as $\hat{\mu} = \hat{\mu}_1 - \hat{\mu}_0$. The equations used in step 2 to estimate parameter $\bflambda$ extract information from the marginal probability of switching under control treatment, i.e. $P(S=1|R=0,\bfC)$ (see appendix \ref{Appendix: identification}).  
As shown in appendix \ref{Appendix: identification}, using these equations to estimate $\bflambda$ leads to a consistent estimator for $\mu$. 
 The variance of $\hat{\mu}$ can  be estimated using the nonparametric bootstrap or 1 over $n$ times the sample variance of the influence function (see appendix \ref{Appendix: influence function}). R code for this estimator is given in appendix \ref{Appendix: implementation in R}.
	 
Even though randomisation is done   independently of the baseline covariates $\bfC$, improvement in the precision of the inverse probability weighted means can be made by using propensity scores $P(R=1|\bfC)$ estimated under more flexible models (Rotnitzky et al., 2010 \cite{rotnitzky2010note}), as explained in appendix \ref{Appendix: identification}.

	\section{Simulations}
	\label{Section: simulations}
	We performed a simulation study to investigate the finite-sample performance of our estimators. The  settings considered are based on the long-term prevention trial in patients who are asymptomatic at risk for developing Alzheimer's dementia. 

	\subsection{Data generation}
	\label{Section: data generation}

 First, we generate the randomised treatment $R \sim \text{Ber}(0.5)$ and baseline covariate $C \sim N(0,1)$. Both treatment groups are assumed to experience a decline in ADCS-PACC scores, with negative mean changes from baseline. The change in score at month 26 is used as variable $L$, indicating how seriously ill the patient is and mainly determining whether the patient needs  rescue medication.
 The potential change in the ADCS-PACC score from baseline at month 26 under treatment is drawn as $L^1 |C \sim N(\delta_1 + \delta_2C, \sigma^2_L)$, for all patients across both arms.
 The distribution of these values is illustrated in figure \ref{fig: scenarios2} in appendix \ref{Appendix: Simulation settings}.  The switching decision for  patients in the treatment arm is generated as $ S| L^1, R=1,C \sim \Bin(\expit(\omega_1+\omega_2C+\omega_3L^1))$, and 
  for  patients in the control arm as  	 $ S | L^1, R=0,C \sim \Bin(\expit(\lambda_1+\lambda_2C+\rho\omega_3L^1))$. The parameter values used  to simulate data are shown in table \ref{table parameter values} in appendix \ref{Appendix: Simulation settings}.	Figure \ref{fig: scenarios} in appendix \ref{Appendix: Simulation settings} illustrates the assumed mean changes in cognitive score during the trial.
 	  The mean changes in scores $L$ of  patients who do not switch to rescue medication follow a linear decline, with a slower decrease for the treated patients compared to  placebo patients. Switchers have  lower $L$ values on average, but their decline in cognitive score slows down after starting rescue medication. 	 
	 Finally, the observed change in the ADCS-PACC score from baseline at month 54 is drawn as $Y|S,L^1,R,C \sim  N( \alpha_1 + \alpha_2S + \alpha_3L^1+\alpha_4C + \alpha_5(1-R), \sigma^2_Y)$. 
	 
	  Three different settings were considered in the simulations. These settings differ in the percentage of patients who switch, the effect of  rescue medication on the cognitive score  and  the strength of the effect of the score at month 26 on the decision to switch.  Two different sample sizes of 200 and 1000 are considered in the simulations. The performance of our estimators was evaluated through a simulation analysis with 5000 runs for each setting and sample size. The standard error of the estimators is calculated as the standard deviation of the estimates. 
	
	\subsection{Results}
	\label{Section: results}
Table \ref{table: Results simulations} summarizes the results of performed simulations for the IPW estimator, with correctly specified sensitivity parameter $\rho = 0.9$. In addition, figure \ref{fig: results simulations} in appendix \ref{Appendix: Simulation results} shows boxplots with the simulation results.  	
All estimators are approximately unbiased. 	
\begin{table}[h]
	\centering
	\renewcommand{\arraystretch}{1}
	\begin{tabular}{llllll}
		\toprule
	 Scenario  & Sample  & Parameter & Bias & SE & Weights  \\ 
	  & size  &  \\ 
		\cmidrule(lr){1-6} 
		 \textbf{Scenario 1} & 200 & $\mu$  & -0.003 & 0.101 & \\ 
		 \multirow{5}{*}{\makecell[l]{ Large treatment effect \\ Limited  switching effect \\ Switchers: \\11\%  in $R=1$ \\24\%  in $R=0$ }}&& $\mu_1$  & -0.003 & 0.080 & 0.437; 1.642 \\ 
		 && $\mu_0$  & 0.001 & 0.064 & \\
		 \cmidrule(lr){2-6} 
		 & 1000 & $\mu$  & 0 & 0.044 &  \\ 
		&& $\mu_1$ &  0 & 0.034 & 0.508; 1.895 \\ 
		 && $\mu_0$  & 0 & 0.028 & \\ 
			\cmidrule(lr){1-6} 
		 \textbf{Scenario 2} & 200 & $\mu$  & -0.021 & 0.122 & \\
		 \multirow{5}{*}{\makecell[l]{ Small treatment effect \\ Limited  switching effect \\ Switchers: \\ 24\%  in $R=1$ \\ 54\%  in $R=0$ }}&& $\mu_1$  &-0.020 & 0.106 & 0.054; 2.132  \\ 
		 && $\mu_0$  & 0.001 & 0.059 & \\ 
		 \cmidrule(lr){2-6}  
		 & 1000 & $\mu$  & -0.004 & 0.058 & \\ 
	     && $\mu_1$ & -0.003 & 0.051 & 0.088; 2.24 \\ 
		 && $\mu_0$  & 0 & 0.026 & \\ 
		 	\cmidrule(lr){1-6} 
		 \textbf{Scenario 3}  & 200 & $\mu$ & -0.062 & 0.164 & \\
		 \multirow{5}{*}{\makecell[l]{ Large treatment effect \\ Large  switching effect \\ Switchers: \\36\%  in $R=1$ \\ 76\%  in $R=0$ }}&& $\mu_1$  & -0.062 & 0.156 & 0.010; 2.398  \\ 
		 && $\mu_0$ & 0.001 & 0.065& \\ 
		 \cmidrule(lr){2-6} 
		 & 1000 & $\mu$  & -0.012 & 0.106 & \\ 
		& & $\mu_1$ & -0.011 & 0.102 &0.023; 2.424   \\ 
		 && $\mu_0$ & 0 & 0.029 &\\ 
		\bottomrule
	\end{tabular}
\caption{\label{Results of the simu}lations investigating the finite sample size performance of the IPW estimator proposed in section \ref{Section: IPW}. The column `weights' shows the 5\% and 95\% percentiles of the weights 
$W_i/n^{-1} \sum_{j = 1}^n R_jW_j  $	among the patients in the experimental treatment arm. }
\label{table: Results simulations}
\end{table}

The estimator for $\mu_1$ has the lowest variance in scenario 1, when  only 17\% of patients across both arms  switch to rescue medication, compared to 39\% in scenario 2  and 56\% in scenario 3. In addition, the dependence between $L^1$ and switching is relatively weaker in the first scenario, as  seen in figure \ref{fig: scenarios} in appendix \ref{Appendix: Simulation settings}. The variance of $\mu$ in the first scenario is similar to the variance of the treatment policy estimand, while the variance of $\mu$ in the other two scenarios is about 4 times the variance of the treatment policy estimand (see table \ref{table: Results simulations different rho} in appendix \ref{Appendix: Simulation results}). 
The skewness in the estimates for $\mu_1$ (see figure \ref{fig: results simulations} in appendix \ref{Appendix: Simulation results}) is caused by weight variability. In particular,  large estimates for $\mu_1$ are caused by a few data simulations where the weight $\WL$ of one treated patient, who switches to rescue medication, becomes so large that it dominates the estimation. Truncation of the weights $\WL$ at the 1\% and 99\% percentiles decreases the variability of the estimates for $\mu_1$ (see  table \ref{table: Results simulations truncated weights} in appendix \ref{Appendix: Simulation results}). However,  this truncation causes the bias to increase.

In table \ref{table: Results simulations different rho} in appendix \ref{Appendix: Simulation results}, we show simulation results for three different values of $\rho$. In particular, we show the results for the IPW estimator if $\rho$ is correctly specified ($\rho = 0.9$) and if $\rho$ is misspecified ($\rho = 0.8$ or 1). Since the differences in the obtained results in terms of bias and SE  are very limited, we find the proposed estimand  not to be  sensitive to the specified value of this sensitivity parameter. 

	\section{Data analysis}
	\label{Section: data analysis}
	
     In this section, we use the balanced estimand in the analysis of a clinical trial, in which  the efficacy and safety of an experimental treatment to improve glycaemic control was tested for patients with type 2 diabetes mellitus (Stenlöf et al., 2013 \cite{stenlof2013efficacy}) (ClinicalTrials.gov
     identifier: NCT01081834). In this 26-week, randomised, stratified, double-blind, placebo-controlled, phase 3 trial, conducted by Janssen Pharmaceuticals, patients (n = 584) received  canagliflozin 100 or 300 mg or placebo (1:1:1) once daily.   Canagliflozin is a sodium glucose co-transporter 2 inhibitor for type 2 diabetes mellitus. 
     Randomisation was stratified according to whether subjects were taking antihyperglycaemic agents (AHA) at screening and whether they participated in the frequently-sampled mixed-meal tolerance test (FS-MMTT).
      The primary endpoint of this study was the change from baseline in haemoglobin A1c (HbA1c) at week 26, comparing both doses of the study medication separately to placebo. Study visits for this primary endpoint and other  secondary endpoints,  were planned at baseline, week 6, week 12, week 18 and week 26. Baseline characteristics are shown in table \ref{table: Baseline characteristics} in appendix \ref{Appendix: Materials and methods}.
     
      During this trial, glycaemic rescue therapy with metformin could be initiated if needed. The decision to take this rescue medication was made by clinicians 
      throughout the study and was mainly based on the observed fasting plasma glucose (FPG)  measurements. 
     A higher percentage of patients treated with placebo (22.7\%) received glycaemic rescue therapy compared to patients treated with canagliflozin 100 and 300 mg (2.6 and 2.0\%). In the primary analysis, the last observation carried forward (LOCF) approach was used to impute the missing efficacy data. In particular, for patients who received rescue therapy, the last post-baseline value prior to initiation of rescue therapy was used.  
     
     Here, we will estimate the treatment effect that would have been observed, had patients on the placebo arm been switched to rescue medication if and only if they would have been switched when randomised to canagliflozin 100 mg, i.e. $E\left(Y^1-Y^{0S^1}\right)$. The switching status is fixed to the value it would have been under treatment and not the value it would have been under placebo, since the large imbalance between the amount of switchers in both groups can cause problems in the identification of $P(S=1|\bfL^1,R=0,\bfC)$ (see appendix \ref{Appendix: identification} for further details). In this setting, the hypothetical estimand may also be relevant since the percentage of switchers is very limited in the canagliflozin 100 mg arm.  In particular, we expect small differences between the balanced estimand and the hypothetical estimand. 
     
     \subsection{Methods}
     
     Efficacy analyses comparing canagliflozin 100 mg to placebo are performed by estimating the  balanced estimand, the treatment policy or ITT estimand and the hypothetical estimand using IPW estimators (see appendices \ref{Appendix: Treatment policy estimand} and \ref{Appendix: Hypothetical estimand}). These estimands can be defined according to the guidelines of the ICH E9(R1) guideline (see appendix \ref{Appendix: Estimands framework data analysis}). The  analysis set consists of all randomised patients who received at least one dose of the study drug. 
     Multiple imputations (MI) using chained equations (Van Buuren et al., 1999 \cite{van1999multiple}) were used to impute the missing efficacy data.  
     We have complete data for 263 patients, meaning  that 124 patients had at least one missing observation (see table \ref{table: Missing data} in appendix \ref{Appendix: Materials and methods}). All baseline covariates except plasma glucose were fully observed. 
       The observed missing data pattern is close to monotone-missingness, but there are also occasional missing values due to patients missing just one study visit.
     Since all variables with missing values are continuous, we used predictive mean matching as imputation method. All observed baseline variables and 5 longitudinal variables (see tables \ref{table: Baseline characteristics} and \ref{table: Missing data} in appendix \ref{Appendix: Materials and methods}), together with the treatment arm, whether or not the patient switches to rescue medication and the study visit at which the patient switches, were used as predictors for the imputation models. These imputations models rely on the MAR assumption. 
     In total, 50 datasets were imputed using 10 iterations. 
     
     Even though patients could initiate rescue medication throughout the study, for this illustration, we assume switching happened at one pre-specified time point. In addition, we summarize the longitudinal FPG and HbA1c values by taking the average before initiation of rescue medication for the switchers and the average before week 26 for the non-switchers and use this as variable $\bfL$ (see figures \ref{fig: results mean HbA1C}, \ref{fig: results time HbA1C} and \ref{fig: results time FPG} in appendix \ref{Appendix: Materials and methods}). The  baseline variables $\bfC$ that were used are shown in table \ref{table: Baseline characteristics} in appendix \ref{Appendix: Materials and methods}. 
     First, the full logistic regression model for $P(S=1|\bfL^0,R=0, \bfC)$, including all variables as main predictors, was fitted in every imputed dataset and pooled using Rubin's rules (Rubin, 2004 \cite{rubin2004multiple}). Next,  variable selection was performed using backward elimination,  where at each step the variable with the highest $p$-value was omitted, the new logistic model was fitted in every dataset and pooled using Rubin's rules. This was terminated when all included predictors had a $p$-value below 10\%.    
      Afterwards, all two-by-two interactions between the variables in the model were each in turn added to the model and kept if the corresponding $p$-value was below 10\%. For every imputed dataset, this results in a model 
       $P(S=1|\bfL^0,R=0, \bfC) = \expit(\omega_1 + \bfomega_2'\bfC + q(\bfL^0,\bfC))$, with $\bfC$ the vector with selected baseline covariates and interactions between them and  $q(\bfL^0,\bfC)$ a linear combination of the  selected longitudinal covariates,  interactions between them and interactions between baseline covariates and  longitudinal covariates.
      Next, $\mu$ is estimated in every imputed dataset. 
      The model for switching in the treatment group was determined as
      $P(S=1|\bfL^0,R=1, \bfC) = \expit(\lambda_1 + \bflambda_2'\bfC + \rho q(\bfL^0,\bfC))$,
       where parameter $\bflambda = (\lambda_1, \bflambda_2)'$ is estimated by solving estimating equations (equations (\ref{Appendix: eq: estimating equation q0 switchers}) in  appendix \ref{Appendix: identification}).  
%
%
     The standard errors are obtained from  1000 nonparametric stratified bootstrap replications of each of the 50 imputed datasets and combined using Rubin's rules. 
       The 95\% confidence intervals are estimated as percentile intervals from the pooled sample of 50$\times$1000 estimates for $\mu$, $\mu_1$ and $\mu_0$. This method is described in Schomaker and Heumann  (2018 \cite{schomaker2018bootstrap}) as the `MI Boot pooled sample method'. In addition, we estimated the standard errors of the balanced estimand using the influence function (IF) (see appendix \ref{Appendix: influence function}).

     \subsection{Results}

	\begin{table}[h]
		\centering
		\begin{tabular}{llr}
			\toprule
			Estimand  & Parameter  & Estimate (95\% CI) \\ 
			\cmidrule(ll){1-3} 
			\multirow{2}{*}{\makecell[l]{Balanced \\ ($\rho = 1$)}} 
			& $E(Y^1-Y^{0S^1})$ & -0.872 [-1.673, -0.494]\\
			& $E(Y^1)$ &   -0.870 [ -0.992, -0.747] \\
			& $E(Y^{0S^1})$ & 0.003 [-0.354,  0.795] \\
			\cmidrule(ll){1-3} 
			\multirow{2}{*}{\makecell[l]{Balanced  \\ ($\rho = 0.9$)}} & $E(Y^1-Y^{0S^1})$ & -0.874 [-1.678, -0.498]\\
			& $E(Y^1)$ & -0.870 [-0.992, -0.748] \\
			& $E(Y^{0S^1})$ & 0.004 [ -0.349, 0.801] \\
			\cmidrule(ll){1-3} 
			\multirow{2}{*}{\makecell[l]{Balanced  \\ ($\rho = 0.8$)}} & $E(Y^1-Y^{0S^1})$ & -0.876 [-1.679, -0.496] \\
			& $E(Y^1)$ & -0.870 [-0.992, -0.748]  \\
			& $E(Y^{0S^1})$ & 0.006 [-0.353,  0.796] \\
			\cmidrule(ll){1-3} 
			Treatment policy & $E(Y^1-Y^{0})$ & -0.642 [-0.832, -0.453]\\
			& $E(Y^1)$ & -0.870 [-0.992, -0.748] \\
			& $E(Y^{0})$ & -0.228 [-0.374, -0.084]\\
		    \cmidrule(ll){1-3} 
			Hypothetical & $E(Y^{10}-Y^{00})$ & -0.897 [-1.529, -0.552]\\
			& $E(Y^{10})$ &-0.875 [-1.003, -0.746] \\
			& $E(Y^{00})$ & 0.022  [-0.294, 0.649]\\
			\bottomrule
		\end{tabular}
		\caption{Results of the data analysis comparing canagliflozin 100 mg to placebo. Estimates for three different estimands regarding the intercurrent event `switching to rescue medication' are shown. 
		}
		\label{table: Results data analysis}
	\end{table}
	 Table \ref{table: Results data analysis} summarizes the results of the data analysis, while the obtained models and  weights  are shown in appendix \ref{Appendix: Results}.  The estimated balanced estimand for $\rho = 0.9$ (-0.874 (bootstrap SE 0.295, IF SE 0.293)), indicates that the decrease from baseline in HbA1c at week 26 would be 0.87\% larger if all patients received canagliflozin 100 mg, compared to placebo, while initiating rescue medication only if they would do that under canagliflozin 100 mg. 
	  As expected, the treatment policy estimand  leads to an attenuated treatment effect (-0.642 (SE 0.097)) compared to the hypothetical estimand (-0.897 (SE 0.251)) and balanced estimand, since a much higher percentage of patients treated with placebo received rescue therapy compared  to patients treated with canagliflozin 100 mg. In addition, the differences between the balanced estimand, where the switching status is fixed to the value it would have been under active treatment,  and the hypothetical estimand are limited, since only 2.6\% of the treated patients initiate rescue medication. Finally, we find the novel estimand not to be  sensitive to the specified value of $\rho$.

	\section{Discussion}
	
	\label{Section: discussion}
	In this article, we have proposed a novel estimand for  the treatment effect in clinical trials where patients can take rescue medication in response to worsening of their disease. 	We defined it  as a natural direct effect and developed a novel inverse probability weighting estimator for it, to account for post-treatment confounding. It is primarily relevant in settings where treatment switches are not uncommon in both arms of the trial. Our proposal relies on untestable assumptions, in view of which we proposed a sensitivity analysis.  Sensitivity analyses for settings with post-treatment confounding  have been described previously in the literature.
	Imai and Yamamoto (2013 \cite{imai2013identification}) proposed a sensitivity analysis to assess the robustness to potential violation of the assumption of no treatment-mediator interactions, 
	using linear structural equation models with random effects to model the outcome and  the mediator of interest.
	However, this approach is not readily applicable to the context of rescue treatment in view of the linearity assumptions on the mediator, and cannot easily accommodate multiple confounders. 
	 Vansteelandt and VanderWeele (2012 \cite{vansteelandt2012natural})  proposed a sensitivity analysis method for post-treatment confounding that involves specifying a selection bias function. This function can be difficult to interpret, but evaluates to zero in a large class of realistic data-generating mechanisms. Their technique, in contrast to ours, requires a model for the density of the possibly high-dimensional confounders $\bfL$.  
	  Concurrent work   (Tchetgen Tchetgen and Shpitser, 2012 \cite{tchetgen2012semiparametric}; VanderWeele and Chiba, 2014 \cite{vanderweele2014sensitivity}) does not assume that data is available on the post-treatment confounder $\bfL$, but  requires  specification of a rather large number of sensitivity parameters.  It is less relevant to our setting where there is a good understanding of what the key confounders are. 

	We used the proposed method for the analysis of a clinical trial conducted by Janssen Pharmaceuticals, in which chronically ill patients can switch to rescue medication for ethical reasons. Application to this clinical trial demonstrated adequate performance.  Monte Carlo simulations confirmed that the proposed estimator is   unbiased in moderate sample sizes. However, in some settings our approach can suffer from weight variability as the weights can become so large that individual observations dominate the estimation. 
	This  instability of IPW estimators in the presence of influential weights is a general concern  (Vansteelandt et al., 2010 \cite{vansteelandt2010analysis}). 
	  In view of this, we plan to develop a more efficient doubly robust estimator that relies on  working models for the outcome and confounders, as well as a model for 
	  the probability to be in the active treatment arm,  
	  but only requires one of these to be correctly specified. 
	   In addition, use of such an estimator will be less sensitive to bias from weight truncation.

	In this paper, we considered the simple setting where patients can switch to rescue medication at one time point during the trial. In future work, we hope to extend our estimand to the setting where the decision to switch can be taken at different times for different patients.  This complicates the estimation of the treatment effect because it requires methods for causal mediation analysis with longitudinal mediators and confounders where problems of post-treatment confounding are even more severe (Daniel et al., 2013 \cite{daniel2013methods}).
	 Likewise, challenges are foreseen to  expand the proposed method to  a censored time-to-event (survival) outcome, in which case one must acknowledge that the event may happen before rescue treatment is initiated.  
	 
	 The proposed estimator for the balanced estimand requires the collection of the outcome after the intercurrent event. However, this is in line with the addendum of the ICH E9 guideline that states that all efforts should be made to collect all data that are relevant to support estimation, including data that inform the characterisation, occurrence and timing of intercurrent events.
	 
	 The novel estimand is not only useful in the considered setting, but 
	  can also be used to handle other intercurrent events such as treatment discontinuation. In that setting, the balanced estimand targets the treatment effect that would have been observed, had patients discontinued treatment if and only if they would have discontinued when randomised to experimental treatment. 		  
	   In addition, the novel IPW estimator can more generally be used to estimate natural direct effects where the association between the mediator and the outcome is confounded by variables which are themselves affected by the exposure.

\bibliographystyle{abbrv}
\bibliography{mybib}
	
	\newpage

	\begin{appendices}
		
		\section{Proposed estimand}
		\label{Appendix: proposed estimators}
		
		\subsection{Estimands framework}
		\label{Appendix: estimands framework}
	The balanced estimand, proposed in section \ref{Section: Natural direct effect}, can be defined according to the estimands framework described in the addendum of the  ICH E9(R1) guideline (International Council for Harmonisation, 2019 \cite{ICH2019}). In the context of the Alzheimer's dementia study, this estimand can be defined using the following attributes: 
		\begin{itemize}
			\item \textbf{Treatment}: experimental treatment or placebo, as defined by the study protocol. 
			\item \textbf{Population}: the entire study population, as defined by the inclusion-exclusion criteria of the study. 
			\item \textbf{Variable}: change from baseline to month 54 in the cognitive endpoint. 
			\item \textbf{Intercurrent events}: switching to rescue medication: the hypothetical scenario is envisaged where patients on the active arm had been switched to rescue medication if and only if they would have been switched when randomised to placebo. 
			\item \textbf{Population-level summary}: difference in means of the variable. 
		\end{itemize}

\subsection{Identification}
\label{Appendix: identification}
In this section, we show how $\mu_1$ can be identified in a much more general way than proposed in section \ref{Section: Identification}.

First, we allow the randomisation to depend on certain baseline covariates, e.g. we do not assume $R \independent \bfC$ here. This ensures, for example, that the proposed method can be applied to a study with stratified randomisation. Consequently, the proposed approach only relies on the following assumptions. 
 The first assumption is the ignorability assumption $Y^{1s} \independent S | \bfL, R = 1, \bfC$ ($s \in \{0,1\}$). To identify $\mu_1$, one also needs to rely on a so-called cross-world independence assumption, i.e. $Y^{1s} \independent S^0| \bfL^1, \bfC$. According to Mittinty and Vansteelandt (2019 \cite{mittinty2019longitudinal}) this can be viewed as a strengthening of the other assumptions needed to identify $\mu_1$. 
  The ignorability assumption expresses that patients who would be switchers versus non-switchers if given active treatment are exchangeable (within strata of $\bfL^1$ and $\bfC$) in terms of what their outcome would be if given active treatment and if the switching status were set to $s \in \{0,1\}$. The cross-worlds independence assumption requires that patients who would be switchers versus non-switchers under control are  exchangeable in terms of what their outcome would be under active treatment and if the switching status were set to $s \in \{0,1\}$.
 In addition, identification of $\mu_1$ relies on 
 randomised assignment conditional on the baseline covariates $\bfC$,  which implies 
 $R \independent Y^{r}|\bfC $ and $R \independent (Y^{1s}, S^0, \bfL^1, \bfC)$ for $ r,s \in \{0,1\}$. 
 
 Second, we assumed the probability $P(S=1|\bfL^1,R=1,\bfC)$ to follow the  parametric model  $P(S=1|\bfL^1,R=1,\bfC) =  \text{expit}(\omega_1 + \bfomega_2'\bfC + \bfomega_3'\bfL^1)$. In this appendix, we  assume  a more general parametric model $\lm$ with unknown parameter value $\bfomega_0$.

Next, instead of assumption (\ref{eq: model q link prob S}), we  assume 
\begin{align}
&P(S=1|\bfL^1,R=0,\mathbf{C})\nn\\
&=P(S=1|\bfL^1,R=1,\mathbf{C}) \frac{\exp\left\{\qzero + \qone\right\}}{E\left[\exp\left\{S(\qzero + \qone)\right\}|\bfL^1,R=1,\mathbf{C}\right]},
\label{Appendix eq: model q link prob S}
\end{align}
with $\qzero$ an unknown function and  $\qone$ a known function that satisfies $q_1(0,\bfC) = 0$ for all $\bfC$.
 Assumption  (\ref{eq: model q link prob S})  corresponds to the choices $\qzero = \lambda_1 - \omega_1 + (\bflambda_2 -\bfomega_2)'\bfC$ and $\qone = \qoneC$ with $\bfomega_3'$ the coefficient of  $\bfL^1$  in the model $\lm = \text{expit}(\omega_1 + \bfomega_2'\bfC + \bfomega_3'\bfL^1)$.   
  For  patients with given severity of disease under treatment $\bfL^1$ and baseline covariates $\bfC$, the probability of switching $S$ in the control group is thus assumed to  equal the probability of switching $S$ in the treatment group tilted by an exponential function.   Tilt functions have previously been proposed for handling nonignorable missing outcome problems (e.g. Scharfstein et al., 1999 \cite{scharfstein1999adjusting}; Vansteelandt et al., 2007 \cite{vansteelandt2007estimation}; Kim and Yu, 2011 \cite{kim2011semiparametric}; Scharfstein et al., 2014 \cite{scharfstein2014global}) and will here be used to address the fact that $\bfL^1$ is missing in the control group. The expectation in the denominator is needed to make sure that the density function $P(S|\bfL^1,R=0,\mathbf{C})$ integrates to 1.

 Assumption (\ref{Appendix eq: model q link prob S}) does not place a restriction on the observed data beyond the restriction $P(S|R=0,\bfC) = \int P(S|\bfL^1,R=0,\bfC)f(\bfL^1|R=1,\bfC)d\bfL^1 $ and that if $P(S = 1|\bfL^1,R=1,\mathbf{C}) = 0$ for certain $\bfL^1$ and $\bfC$, $P(S = 1|\bfL^1,R=0,\mathbf{C})$ is also 0.  In particular, from  
 \begin{align*}
 &P(S = 0|R = 0,\bfC) \\ 
 &= \int P(S = 0| \bfL^1,R=0,\bfC) f(\bfL^1|R=0,\bfC)d\bfL^1 \\
 &= \int  P(S = 0| \bfL^1,R=1,\bfC) \frac{1}{E\left[\exp\left\{S(\qzero + \qone)\right\}|\bfL^1,R=1,\mathbf{C}\right]} f(\bfL^1|R=1,\bfC)d\bfL^1 \\
 &= P(S = 0|R = 1,\bfC) E\left[\frac{1}{E\left[\exp\left\{S(\qzero + \qone)\right\}|\bfL^1,R=1,\mathbf{C}\right]} |S=0,R=1,\bfC\right],
 \end{align*}
 where we use that $\bfL^1 \independent R |\bfC$, 
 it follows that 
 \begin{eqnarray}
 	&&E\left[\frac{(1-R)(1-S)}{P(R = 0|\bfC)} - \frac{R(1-S)}{P(R = 1|\bfC)}  \frac{1}{E\left[\exp\left\{S(\qzero + \qone)\right\}|\bfL^1,R=1,\mathbf{C}\right]} \right] \label{Appendix: eq: estimating equation q0} \\
 	&&= E\left[P(S = 0|R=0,\bfC) \right. \nonumber \\
 	&& \left. - E\left(\frac{1}{E\left[\exp\left\{S(\qzero + \qone)\right\}|\bfL^1,R=1,\mathbf{C}\right]} |S = 0,R = 1,\bfC \right)P(S = 0|R = 1,\bfC) \right] \nonumber\\
 	&&= 0. \nonumber
 \end{eqnarray}
Therefore,  the  unknown parameter value $\bflambda_0$, indexing model ${q_0}(\bfC;{\bflambda})$, can be consistently estimated by solving  estimating equation (\ref{Appendix: eq: estimating equation q0}). Since the densities $P(R|\bfC)$ and $P(S|\bfL^1,R=1,\bfC)$ can be obtained from the observed data, it can be
inferred from this that $\bflambda$ is identifiable from the observed data when $\qone$ is given. 
Since restriction (\ref{Appendix: eq: estimating equation q0}) is the only testable restriction on the observed data it further follows that $\qone$ is not identifiable from the observed data when $\qzero$ is left unspecified, and thus that each choice of $\qone$ is compatible with the observed data. 
 After specifying $\qone$ and fitting 
 a model $\pi(\bfC;\hat{\bftheta})$ for the probability $P(R=1|\bfC)$ and a model $l(\bfL,\bfC;\hat{\bfomega})$ for  $P(S=1|\bfL,R=1,\bfC)$, the  unknown parameter value $\bflambda_0$ indexing model ${q_0}(\bfC;{\bflambda})$ can be estimated by solving 
 \begin{eqnarray}
 	&&\mathbf{0} = \sum_{i = 1}^n \pmb{\psi}(\bfC_i) \left( \frac{(1-R_i)(1-S_i)}{1-\pi(\bfC_i;\hat{\bftheta})} \right.  \nn\\ 	&& \left.-\frac{R_i(1-S_i)}{\pi(\bfC_i;\hat{\bftheta})}  \frac{1}
 	{l(\bfL_i,\bfC_i;\hat{\bfomega})\left\{\exp\left({q_0}(\bfC_i;\hat{\bflambda}) + {q_1}(\bfL_i,\bfC_i) \right)-1 \right\} +1 }\right) \label{Appendix: eq: estimating equation q0 nonswitchers}.
 \end{eqnarray}
 In this equation, $\pmb{\psi}$ is an arbitrary function of $\bfC$ of the same dimension as $\bflambda$. For instance, when e.g. ${q_0}(\bfC;{\bflambda}) = \lambda_1 +  \bflambda_2'\bfC$, $\pmb{\psi}(\bfC) = (1,\bfC)'$ can be used. 
Note that the data of the switchers is not used in this equation. Similarly, it can be shown that $\bflambda$ can be estimated by solving estimating equations
\begin{eqnarray}
	&&\mathbf{0} = \sum_{i = 1}^n \pmb{\psi}(\bfC_i) \left( \frac{(1-R_i)S_i}{1-\pi(\bfC_i;\hat{\bftheta})} \right. \nn \\ 	&& \left.-\frac{R_iS_i}{\pi(\bfC_i;\hat{\bftheta})}  \frac{\exp\left({q_0}(\bfC_i;\hat{\bflambda}) + {q_1}(\bfL_i,\bfC_i) \right)}
	{l(\bfL_i,\bfC_i;\hat{\bfomega})\left\{\exp\left({q_0}(\bfC_i;\hat{\bflambda}) + {q_1}(\bfL_i,\bfC_i) \right)-1 \right\} +1 }\right), \label{Appendix: eq: estimating equation q0 switchers}
\end{eqnarray}
using of the observed data of the switchers instead of the non-switchers. Depending on the context equation (\ref{Appendix: eq: estimating equation q0 nonswitchers}) or  (\ref{Appendix: eq: estimating equation q0 switchers}) can lead to more accurate estimates for $\bflambda$. 
  
 The weights $\WL$ can next be estimated as  
 \begin{align*}
 	W(S,\bfL,\bfC; \hat{\bflambda}, \hat{\bfomega} ) &= \frac{\exp(S({q_0}(\bfC;\hat{\bflambda}) + \qoneL))}{E[\exp(S({q_0}(\bfC;\hat{\bflambda}) + \qoneL))|\bfL,R=1,\bfC ]} \\
 	&= \frac{\exp(S({q_0}(\bfC;\hat{\bflambda}) + \qoneL))}{l(\bfL,\bfC;\hat{\bfomega})\left\{\exp\left({q_0}(\bfC;\hat{\bflambda}) + {q_1}(\bfL,\bfC) \right)-1 \right\} +1} .
 \end{align*}
 
 Even when randomisation is done   independently of the baseline covariates $\bfC$, improvement in the precision of the inverse probability weighted means can be made by using propensity scores $P(R=1|\bfC)$ estimated under more flexible models (Rotnitzky et al., 2010 \cite{rotnitzky2010note}).
 Therefore, it is in principle preferable to  fit a parametric model $\pi(\bfC;\pmb{\theta})$ for the probability of assignment to the active treatment arm $P(R=1|\bfC)$; e.g. $\pi(\bfC;\pmb{\theta}) = \text{expit}(\theta_1 + \pmb{\theta}_2'\bfC)$ and replace $\hat{\pi}$ by $\pi(\bfC;\hat{\pmb{\theta}})$ in step 2 of the IPW approach in section \ref{Section: IPW}. Parameter $\mu_1$ can then be estimated as 	$\hat{\mu}_1 =  \sum_{i = 1}^n Y_i	\frac{{R_i}W_i}{{\pi}(\bfC_i,\hat{\bftheta})} / \sum_{i = 1}^n 	\frac{{R_i}W_i}{{\pi}(\bfC_i,\hat{\bftheta})}$
  and $\mu_0$ as $\hat{\mu}_0 =  \sum_{i = 1}^n Y_i\frac{1-R_i}{1-{\pi}(\bfC_i,\hat{\bftheta})} / \sum_{i = 1}^n \frac{1-R_i}{1-{\pi}(\bfC_i,\hat{\bftheta})}$.

 The restriction of assumption (\ref{Appendix eq: model q link prob S}) which states that if $P(S = 1|\bfL^1,R=1,\mathbf{C}) = 0$ for certain $\bfL^1$ and $\bfC$, $P(S = 1|\bfL^1,R=0,\mathbf{C})$ is also 0, might be  violated when there is a large imbalance in switching events in both groups. This is e.g. the case when nobody would change to rescue medication when assigned to the experimental treatment arm of the study, i.e. if there are no switchers under treatment, but some patients would change to rescue medication when assigned to control. 
 In that setting, it might better to target the balanced estimand where  the switching status is fixed to the natural value it would have been under experimental treatment: $E\left(Y^{1}-Y^{0S^1} \right)$.  This effect can simply be estimated using the proposed IPW method, upon interchanging the meaning of variable $R=0$ and $R=1$ in the dataset. 

The sum of $\qzero$ and $\qone$ can be interpreted as the log odds ratio for switching for the control group versus the treatment group: 
\begin{align}
\exp\{\qzero+\qone\} &= \frac{\odds(S=1|R=0, \bfL^1,\bfC)}{\odds(S=1|R=1, \bfL^1,\bfC)}. 
\end{align}
Here, $\qone$ represents the extent to which the association between switching and severity of disease under treatments differs between the two treatment groups: 
\begin{align}
\exp\{\qone\} &= \frac{\odds(S=1|R=0, \bfL^1,\bfC)/\odds(S=1|R=1, \bfL^1,\bfC)}{\odds(S=1|R=0, \bfL^1=\mathbf{0},\bfC)/\odds(S=1|R=1, \bfL^1=\mathbf{0},\bfC)}.
\label{Appendix eq: meaning q1}
\end{align}
The value $\bfL^1 = \mathbf{0}$ is a reference value, e.g. not severely ill under treatment. 

The function $\qone$ needs to be specified by the user since $\bfL^1$ is not observed for the control patients.  
In this section, we  discuss an extreme value for  $\qone$ and afterwards, we  propose a sensitivity analysis. 
From (\ref{Appendix eq: meaning q1}), it follows that the choice $\qone \equiv 0$ implies 
\begin{equation*}
\frac{\odds(S=1|R=0, \bfL^1,\bfC)}{\odds(S=1|R=1, \bfL^1,\bfC)} = \frac{\odds(S=1|R=0, \bfL^1=\mathbf{0},\bfC)}{\odds(S=1|R=1, \bfL^1=\mathbf{0},\bfC)},
\end{equation*}
meaning that differences between the switching statuses in both treatment groups can be fully assigned to the baseline covariates. This is generally implausible. 
In particular, if  patients are more seriously ill under control than under treatment, we expect odds ratios in the numerator and denominator of (\ref{Appendix eq: meaning q1}) to be larger than one. The odds ratio in the denominator is the odds ratio for patients who would not have been seriously ill under treatment, e.g. $\bfL^1= \mathbf{0}$. 
Since switching $S$ is more weakly associated with $\bfL^1$ in the control group than in the treatment group, we expect $\qone$ to be smaller than 0.   

\subsection{Influence function}
\label{Appendix: influence function}

 In this section, we derive the influence function of the proposed IPW estimator of the parameter $\mu$.  First, we derive the influence function of $\hat{\mu}_{1}$.

Since 
\begin{equation*}
\hat{\mu}_{1} = \left\{\sum_{j=1}^n R_j W(S_j,\bfL_j,\bfC_j;\hat{\bflambda}, \hat{\bfomega})/\pi(\bfC_j;\hat{\bftheta}) \right\}^{-1} \sum_{i=1}^n Y_i R_i W(S_i,\bfL_i,\bfC_i;\hat{\bflambda}, \hat{\bfomega})/\pi(\bfC_i;\hat{\bftheta}), 
\end{equation*}
it holds that 
\begin{eqnarray*}
	0 = \frac{1}{\sqrt{n}}\sum_{i = 1}^{n} R_i W(S_i,\bfL_i,\bfC_i;\hat{\bflambda}, \hat{\bfomega})/\pi(\bfC_i;\hat{\bftheta})(Y_i-\hat{\mu}_1). 
\end{eqnarray*}
From the Taylor expansion around $\mu_1$, $\bflambda$, $\bfomega$ and $\bftheta$ and the uniform WLLN (see Newey and McFadden (1994 \cite{newey1994large}), Lemma 4.3), we obtain 
\begin{eqnarray}
	0 &&= \frac{1}{\sqrt{n}} \sumYi - \sqrt{n} \mu_1 \frac{1}{n} \sumi \nn \\
	&&+  \frac{1}{n} \sum_{i=1}^n R_i \Wderivi (Y_i - \mu_1) \sqrt{n} \lambdaomega \nn \\
	&&- \frac{1}{n} \sumi \sqrt{n} (\hat{\mu}_1-\mu_1) + \op  \nn\\
	&&=  \frac{1}{\sqrt{n}} \sumYi - \sqrt{n} \mu_1 \nn \\
	&&+  \frac{1}{n} \sum_{i=1}^n R_i \Wderivi (Y_i - \mu_1) \sqrt{n} \lambdaomega \nn \\
	&&-  \sqrt{n} (\hat{\mu}_1-\mu_1) + \op,
	\label{eq: IF mu1}
\end{eqnarray}
where $\Wi \equiv \Wilong$ and $\pi_i(\bftheta)\equiv \pi(\bfC_i;\bftheta)$.
Since parameter $\bfomega$ indexing 
\begin{equation*}
\lmL = \expit(\omega_1+\bfomega'_2\bfC+\bfomega'_3\bfL)=  P(S=1|\bfL,R=1,\bfC)
\end{equation*}
is unknown and substituted by the MLE, we have 
\begin{equation}
\sqrt{n}(\hat{\bfomega} - \bfomega) = \frac{1}{\sqrt{n}} \sum_{i=1}^n E(\bfS_{i,\bfomega}\bfS_{i,\bfomega}')^{-1}\bfS_{i,\bfomega} +\op.
\label{eq: asymptotic beh omega}
\end{equation}
Here, $\bfS_{i,\bfomega}$ denotes the score vector for $\bfomega$ in  individual $i$. 

Parameter $\bflambda = (\lambda_1,\bflambda_2)'$ indexing $P(S = 1|\bfL^1,R=0,\mathbf{C}) = \expit(\lambda_1 + \bflambda'_2\bfC + \rho\bfomega'_3\bfL^1)$ is estimated by solving estimating equations. In particular, the estimator $\hat{\bflambda}$ for the parameter $\bflambda$ is the solution to the estimating equations $E(\bfU_\bflambda(S,\bfL,R,\bfC;\hat{\bflambda}, \hat{\bftheta})) = \mathbf{0}$ with 
\begin{eqnarray*}
	&&\bfU_\bflambda(S,\bfL,R,\bfC;{\bflambda}, {\bftheta}) = \begin{pmatrix}
		1 \\ \bfC
	\end{pmatrix} \left( \frac{(1-R)S}{1-\pi(\bfC;{\bftheta})} -\frac{RS}{\pi(\bfC;{\bftheta})}  \frac{\exp\left({q_0}(\bfC;{\bflambda}) + {q_1}(\bfL,\bfC) \right)}
	{l(\bfL,\bfC;{\bfomega})\left\{\exp\left({q_0}(\bfC;{\bflambda}) + {q_1}(\bfL,\bfC) \right)-1 \right\} +1 }\right) \label{eq: estimating equation q0 switchers}.
\end{eqnarray*}
Here, $\qzero = \lambda_1 - \omega_1 + (\bflambda_2 -\bfomega_2)'\bfC$ and $\qoneL = \qoneLC$. Parameter $\bftheta = (\theta_1,\bftheta_2)'$,  indexing $P(R=1|\bfC) = \expit(\theta_1+\bftheta_2'\bfC)$, is unknown and substituted by the MLE $\hat{\bftheta}$. Therefore, this estimator is the solution to the estimating equations $E(\bfU_\bftheta(R,\bfC; \hat{\bftheta})) = \mathbf{0}$ with 
\begin{eqnarray*}
	\bfU_\bftheta(R,\bfC;{\bftheta}) = \begin{pmatrix}
		1 \\ \bfC
	\end{pmatrix} (R -  \expit(\theta_1+\bftheta_2'\bfC)).
\end{eqnarray*}
In addition, we have 
\begin{equation}
\sqrt{n}(\hat{\bftheta} - \bftheta) = \frac{1}{\sqrt{n}} \sum_{i=1}^n E(\bfS_{i,\bftheta}\bfS_{i,\bftheta}')^{-1}\bfS_{i,\bftheta} +\op.
\label{eq: asymptotic beh theta}
\end{equation}
Here, $\bfS_{i,\bftheta} = \bfU_\bftheta(R_i,\bfC_i;{\bftheta})$ denotes the score vector for $\bftheta$ in individual $i$. 
Consequently, the influence function for $\bflambda$ is given by 
\begin{eqnarray*}
	&&\phi_\bflambda(S,\bfL,R,\bfC) \\
	&&=  \bfU_\bflambda - E\left(\frac{\partial \bfU_\bflambda }{\partial \bftheta} \right) E^{-1}\left(\frac{\partial \bfU_\bftheta }{\partial \bftheta} \right) \bfU_\bftheta \\
	&&= \begin{pmatrix}
		1 \\ \bfC
	\end{pmatrix} \left( \frac{(1-R)S}{1-\expit(\theta_1+\bftheta'_2\bfC)} -\frac{RS}{\expit(\theta_1+\bftheta'_2\bfC)}  \frac{\exp\left({q_0}(\bfC;{\bflambda}) + {q_1}(\bfL,\bfC) \right)}
	{l(\bfL,\bfC;{\bfomega})\left\{\exp\left({q_0}(\bfC;{\bflambda}) + {q_1}(\bfL,\bfC) \right)-1 \right\} +1 }\right) \\
	&&+ E\left(\begin{pmatrix}
		1 & \bfC' \\ \bfC  & \bfC\bfC'
	\end{pmatrix}\left[(1-R)S(\exp(\theta_1+\bftheta'_2\bfC)+1) \right. \right. \\
	&& \left. \left.
	-RS(\exp(-(\theta_1+\bftheta'_2\bfC))+1)   \frac{\exp\left({q_0}(\bfC;{\bflambda}) + {q_1}(\bfL,\bfC) \right)}
	{l(\bfL,\bfC;{\bfomega})\left\{\exp\left({q_0}(\bfC;{\bflambda}) + {q_1}(\bfL,\bfC) \right)-1 \right\} +1 }\right] \right) \\
	&& \times E^{-1}\left(\begin{pmatrix}
		1 & \bfC' \\ \bfC  & \bfC\bfC'
	\end{pmatrix} \frac{\exp(\theta_1 + \bftheta'_2\bfC)}{(1+\exp(\theta_1 + \bftheta'_2\bfC))^2}  \right) \begin{pmatrix}
		1 \\ \bfC
	\end{pmatrix} (R-\expit(\theta_1 + \bftheta'_2\bfC)). 
\end{eqnarray*}
As a  consequence, we obtain
\begin{equation}
\sqrt{n}(\hat{\bflambda}-\bflambda) = \frac{1}{\sqrt{n}} \sum_{i=1}^n \phi_\bflambda(S_i,\bfL_i,R_i,\bfC_i) +\op. 
\label{eq: asymptotic beh lambda}
\end{equation}

Using (\ref{eq: IF mu1}), (\ref{eq: asymptotic beh omega}), (\ref{eq: asymptotic beh theta}), (\ref{eq: asymptotic beh lambda}) and the uniform WLLN it follows that
\begin{eqnarray}
&&\sqrt{n}(\hat{\mu}_{1} - \mu_1) \nn \\
&&= \frac{1}{\sqrt{n}} \sum_{i=1}^n Y_i R_i W_i({\bflambda},{\bfomega})/\pii   - \sqrt{n} E[Y_i^{1S_i^0}] \nn \\
&&+   \frac{1}{\sqrt{n}} E\left[Y_iR_i\Wderivi
\right] \sum_{k=1}^n \begin{pmatrix}
\phi_\bflambda(S_k,\bfL_k,R_k,\bfC_k) \\ E(\bfS_{k,\bfomega}\bfS_{k,\bfomega}')^{-1}\bfS_{k,\bfomega} \\ E(\bfS_{k,\bftheta}\bfS_{k,\bftheta}')^{-1}\bfS_{k,\bftheta}
\end{pmatrix} \nn \\
&& - \frac{1}{\sqrt{n}} E\left[R_i \Wderivi \right]  \sum_{k=1}^n \begin{pmatrix}
\phi_\bflambda(S_k,\bfL_k,R_k,\bfC_k) \\ E(\bfS_{k,\bfomega}\bfS_{k,\bfomega}')^{-1}\bfS_{k,\bfomega} \\ E(\bfS_{k,\bftheta}\bfS_{k,\bftheta}')^{-1}\bfS_{k,\bftheta}
\end{pmatrix} \nn \\
&&\times  E[Y_iR_iW_i(\bflambda,\bfomega)/\pii] + \op. \nn 
\end{eqnarray}
We conclude that $\hat{\mu}_1$ is an asymptotically linear estimator of $\mu_1$ with influence function 
\begin{eqnarray}
&&\phi_{\mu_1}(Y_i,S_i,\bfL_i,R_i,\bfC_i; \bflambda, \bfomega) \nn\\
&&= - E[Y_i^{1S_i^0}] +   Y_i R_i W_i({\bflambda},{\bfomega})/\pii \nn \\
&& +  E\left[Y_iR_i \Wderivi \right] \begin{pmatrix}
\phi_\bflambda(S_i,\bfL_i,R_i,\bfC_i) \\ E(\bfS_{i,\bfomega}\bfS_{i,\bfomega}')^{-1}\bfS_{i,\bfomega} \\ E(\bfS_{i,\bftheta}\bfS_{i,\bftheta}')^{-1}\bfS_{i,\bftheta}
\end{pmatrix}\nn \\
&& 
-E\left[R_i \Wderivi \right] \begin{pmatrix}
\phi_\bflambda(S_i,\bfL_i,R_i,\bfC_i) \\ E(\bfS_{i,\bfomega}\bfS_{i,\bfomega}')^{-1}\bfS_{i,\bfomega} \\ E(\bfS_{i,\bftheta}\bfS_{i,\bftheta}')^{-1}\bfS_{i,\bftheta}
\end{pmatrix} \nn \\
&&\times  E[Y_iR_iW_i(\bflambda,\bfomega)/\pii].
\end{eqnarray}
Therefore, the asymptotic variance of $\hat{\mu}_1$ can be calculated as one over $n$ times the sample variance of $\phi_{\mu_1}(Y,S,\bfL,R,\bfC; \hat{\bflambda}, \hat{\bfomega})$.

Since 
\begin{eqnarray*}
	\Wj &&= \left(\frac{\expit(\Tj)}{\expit(\Nj)}  \right)^{S_j}\left(\frac{1-\expit(\Tj)}{1-\expit(\Nj)}  \right)^{1-S_j} \\
	\pij &&= \expit(\theta_1+\bftheta'_2\bfC_j), 
\end{eqnarray*}
it holds that 
\begin{eqnarray*}
	&&\frac{\partial \Wj/\pij }{\partial \bflambda} \\
	&&= \frac{1}{\pij} \frac{\expit(\Tj)}{1+\exp(\Tj)} \left(\frac{1}{\expit(\Nj)}\right)^{S_j} \\ 
	&& \times \left(-\frac{1}{1-\expit(\Nj)}\right)^{1-S_j} \begin{pmatrix}
		1 & \bfC'_j
	\end{pmatrix},
\end{eqnarray*}

\begin{eqnarray*}
	\frac{\partial \Wj/\pij }{\partial \bfomega} = \begin{pmatrix}
		\frac{\partial \Wj /\pij}{\partial \omega_1} & 	\frac{\partial \Wj/\pij }{\partial \omega_1} \bfC'_j & 	\frac{\partial \Wj/\pij }{\partial \bfomega_3}
	\end{pmatrix},
\end{eqnarray*}
with 
\begin{eqnarray*}
	&& \frac{\partial \Wj/\pij }{\partial \omega_1}   \\
	&&= \frac{1}{\pij} \left(-\frac{\expit(\Tj)}{\exp(\Nj)}\right)^{S_j} \\
	&& \times\left((1-\expit(\Tj)\exp(\Nj) \right)^{1-S_j} \\
	&&\frac{\partial \Wj/\pij }{\partial \bfomega_3} \\
	&&=  \bfL_j \frac{1}{\pij} \left(\frac{\frac{\rho\expit(\Tj)}{1+\exp(\Tj)} - \frac{\expit(\Tj)}{1+\exp(\Nj)}}{\expit(\Nj)}\right)^{S_j} \\
	&&\times \left(\frac{\frac{-\rho\expit(\Tj)}{1+\exp(\Tj)} + (1-\expit(\Tj))\expit(\Nj)}{1-\expit(\Nj)}\right)^{1-S_j},
\end{eqnarray*}
and 
\begin{eqnarray*}
	\frac{\partial \Wj/\pij }{\partial \bftheta} = -\Wj \exp(-(\theta_1+\bftheta'_2\bfC_j))\begin{pmatrix}
		1 &\bfC_j'              
	\end{pmatrix}.
\end{eqnarray*}

Next, we derive the influence function of $\hat{\mu}_0$. Since 
\begin{eqnarray*}
	\hat{\mu}_0 &&= \left\{\sum_{j=1}^n (1-R_j)/(1-\pi(\bfC_j;\hat{\bftheta})) \right\}^{-1}  \sum_{i=1}^n Y_i(1-R_i)/(1-\pi(\bfC_i;\hat{\bftheta})),
\end{eqnarray*}
it holds that 
\begin{eqnarray*}
0 	&&= \frac{1}{\sqrt{n}} \sum_{i=1}^n \frac{1-R_i}{1-\pi(\bfC_i;\hat{\bftheta})} (Y_i - \hat{\mu}_0). 
\end{eqnarray*}
From the Taylor expansion around $\mu_0$ and $\bftheta$ and the uniform WLLN, we obtain 
\begin{eqnarray*}
	0 	
	&&=  \frac{1}{\sqrt{n}} \sum_{i=1}^n \frac{1-R_i}{1-\pii} (Y_i - {\mu}_0) + \frac{1}{\sqrt{n}} \sum_{i=1}^n(1-R_i) \frac{\partial}{\partial \bftheta}\left(\frac{1}{1-\pii} \right) (Y_i - {\mu}_0) \begin{pmatrix}
		\hat{\bftheta} - \bftheta
	\end{pmatrix} \\
	&& - \frac{1}{\sqrt{n}} \sum_{i=1}^n \frac{1-R_i}{1-\pii}  (
		\hat{\mu}_0 - \mu_0) + \op \\
	&&= \frac{1}{\sqrt{n}} \sum_{i=1}^n \frac{1-R_i}{1-\pii} Y_i - \sqrt{n} \mu_0 \\
	&&+ \frac{1}{{n}} \sum_{i=1}^n (1-R_i) \frac{\partial}{\partial \bftheta}\left(\frac{1}{1-\pii}   \right) (Y_i - {\mu}_0) \sqrt{n}  \begin{pmatrix}
		\hat{\bftheta} - \bftheta
	\end{pmatrix} \\
	&& - \sqrt{n}(\hat{\mu}_0-\mu_0) + \op. 
\end{eqnarray*}

Consequently, from (\ref{eq: asymptotic beh theta}) and the uniform WLLN it follows that 
\begin{eqnarray*}
	&&\sqrt{n}(\hat{\mu}_0- \mu_0) \nn \\
	&&= \frac{1}{\sqrt n}  \sum_{i=1}^n Y_i(1-R_i)/(1-\pii)  \\
	&&+ E\left[Y_j(1-R_j)\exp(\theta_1+\bftheta'_2\bfC_j)\begin{pmatrix}
		1 & \bfC_j
	\end{pmatrix} \right] \frac{1}{\sqrt n} \sum_{k=1}^n
	E(\bfS_{k,\bftheta}\bfS_{k,\bftheta}')^{-1}\bfS_{k,\bftheta} \nn \\
	&&  - E\left[(1-R_j) \exp(\theta_1+\bftheta'_2\bfC_j)  \begin{pmatrix}
		1 & \bfC_j
	\end{pmatrix} \right] E[Y_j(1-R_j)/(1-\pij)] \frac{1}{\sqrt n} \sum_{k=1}^n
	E(\bfS_{k,\bftheta}\bfS_{k,\bftheta}')^{-1}\bfS_{k,\bftheta}  \nn \\
	&& +\op  -\sqrt{n}E(Y_j^0).
\end{eqnarray*}
As a consequence, $\hat{\mu}_0$ is an asymptotically linear estimator of $\mu_0$ with influence function 
\begin{eqnarray}
&&\phi_{\mu_0}(Y_i,R_i,\bfC_i; \bftheta) \nn\\
&&=  -E(Y_i^0) +  Y_i(1-R_i)/(1-\pii) \nn  \\
&&+ E\left[Y_i(1-R_i)\exp(\theta_1+\bftheta'_2\bfC_i)\begin{pmatrix}
1 & \bfC_i
\end{pmatrix} \right] 
E(\bfS_{i,\bftheta}\bfS_{i,\bftheta}')^{-1}\bfS_{i,\bftheta} \nn \\
&&  - E\left[(1-R_i) \exp(\theta_1+\bftheta'_2\bfC_i)  \begin{pmatrix}
1 & \bfC_i
\end{pmatrix} \right] E[Y_i(1-R_i)/(1-\pii)] 
E(\bfS_{i,\bftheta}\bfS_{i,\bftheta}')^{-1}\bfS_{i,\bftheta}.
\label{eq: IF mu0}
\end{eqnarray}
From (\ref{eq: IF mu1}) and (\ref{eq: IF mu0}), it follows that  $\hat{\mu}$ is an asymptotically linear estimator of $\mu$ with influence function $\phi_{\mu}(Y,S,\bfL,R,\bfC; \bflambda, \bfomega, \bftheta) = \phi_{\mu_1}(Y,S,\bfL,R,\bfC; \bflambda, \bfomega) -\phi_{\mu_0}(Y,R,\bfC; \bftheta)$. The asymptotic variance of $\hat{\mu}$ can thus be calculated as one over $n$ times the sample variance of $\phi_{\mu}(Y,S,\bfL,R,\bfC; \hat{\bflambda}, \hat{\bfomega}, \hat{\bftheta})$.

\subsection{Sensitivity analysis}
\label{Appendix: sensitivity analysis}
	
	In this section, we show a data generating mechanism under which the choice of $\qone = (\rho-1) \bfomega_3'\bfL^1$, proposed in the main paper, is justified. However, the settings under which this choice of $\qone$ is correct are not limited to the settings discussed here.

	 We consider univariate variables $L^0$ and $L^1$ that are normally distributed conditional on the baseline covariates. In addition, we define the correlation $\rho := \text{corr}(L^0,L^1|\bfC)$ between the health statuses under both treatments. 	This correlation cannot be observed and will be used as a sensitivity parameter. However, we expect a rather high correlation between both health statuses. It follows that $L^0|L^1,\bfC$ is also normally distributed with mean 
		\[E(L^0|L^1,\bfC) = E(L^0|\bfC) + \rho (L^1-E(L^1|\bfC))\frac{SD(L^0|\bfC)}{SD(L^1|\bfC)}\]
		and variance 
		\[\text{var}(L^0|L^1,\bfC) = \text{var}(L^0|\bfC)(1-\rho^2). \]
		
		Suppose model $\qzero$ is linear in $\bfC$, i.e. ${q_0}(\bfC;{\bfzeta}) = \zeta_1 +  \bfzeta_2'\bfC$, and $\qonel$ is linear in $L^1$, i.e. ${q_1}(L^1) = \alpha L^1$ for some parameter value $\alpha$. Using (\ref{Appendix eq: model q link prob S}), this leads to the following equality:  
		\begin{equation*}
		P(S=1|L^1,R=0,\bfC)=\frac{P(S=1|L^1,R=1,\bfC)\exp\left(\zeta_1 +  \bfzeta_2'\bfC + \alpha L^1\right)}
		{P(S=1|L^1,R=1,\bfC)\left\{\exp\left(\zeta_1 +  \bfzeta_2'\bfC + \alpha L^1\right)-1 \right\} +1 }.
		\end{equation*}
		If the probability  $P(S=1|L,R,\bfC)$ follows a logistic regression model, i.e. 
		\begin{equation}
		P(S=1|L,R,\bfC) = \expit(\omega_1+ \bfomega_2'\bfC +\omega_3L + \omega_4R ),
		\label{eq: switching prob}
		\end{equation}
		these choices for $\qzero$ and $\qonel$ imply that the probability to switch in the control group conditional on $L^1$ also follows a logistic regression model: 
		\begin{align*}
		&P(S=1|L^1,R=0,\bfC) \nonumber \\
		&= \expit(\omega_1+ \bfomega_2'\bfC +\omega_3L+ \omega_4 ) \frac{\exp(\zeta_1 +  \bfzeta_2'\bfC + \alpha L^1 ) }{\expit(\omega_1+ \bfomega_2'\bfC +\omega_3L + \omega_4 )\left\{\exp\left(\zeta_1 +  \bfzeta_2'\bfC + \alpha L^1\right)-1 \right\} +1 } \nonumber \\
		&=\expit(\omega_1 + \omega_4 + \zeta_1  + (\bfomega_2 + \bfzeta_2)'\bfC + (\alpha + \omega_3)L^1).
		\end{align*}
		The assumed model in section \ref{Section: Identification} of the main paper corresponds to the choices $\lambda_1$ = $\omega_1 + \omega_4 + \zeta_1$ and $\bflambda_2$ = $\bfomega_2 + \bfzeta_2$. 		If switching is rare, this probability can be approximated by an exponential function: 
		\begin{align}
		P(S=1|L^1,R=0,\bfC) \approx \exp(\omega_1 + \omega_4 + \zeta_1 + (\bfomega_2 + \bfzeta_2)'\bfC + (\alpha + \omega_3)L^1 ).
		\label{eq: unobserved prob 1}
		\end{align}
		
		From (\ref{eq: switching prob}), it also follows that 
		\begin{align*}
		P(S=1|L^1,R=0,\bfC) &= \int P(S=1|L^1,L^0, R=0,\bfC) f(L^0|L^1, R=0,\bfC) dL^0 \\
		&= E\left[P(S=1|L^0,R=0,\bfC) |L^1,R=0,\bfC \right] \\
		&= E\left[\expit(\omega_1 + \bfomega_2'\bfC +\omega_3L^0 ) |L^1,R=0,\bfC \right], 
		\end{align*}
		where the second equality follows from the assumption that if $L^0$ is given,  the decision to switch is independent of $L^1$ for a patient in the control group:  $S \independent L^1 |L^0,R=0,\bfC$. This is a reasonable assumption since the decision to take rescue medication is based on $L^0$ and $\bfC$ for a patient in the control group. 
		If switching is  rare, this probability can be approximated: 
		\begin{align*}
		P(S=1|L^1,R=0,\bfC) &\approx E\left[\exp(\omega_1+ \bfomega_2'\bfC +\omega_3L^0 ) |L^1,R=0,\bfC \right]. 
		\end{align*} 
		Using the moment-generating function of a normally distributed variable, this leads to the
		following approximation for the probability to switch in the control group: 
		\begin{align}
		&P(S=1|L^1,R=0,\bfC) \nonumber \\
		&\approx \exp(\omega_1 + \bfomega_2'\bfC ) E\left[\exp(\omega_3L^0 ) |L^1,R=0,\bfC \right] \nonumber \\
		&= \exp\left(\omega_1 + \rho\omega_3L^1 + \bfomega_2'\bfC +\omega_3(E(L^0|\bfC)-\rho E(L^1|\bfC) +  \frac{1}{2}\text{var}(L^0|\bfC)(1-\rho^2) \omega_3  ) \right),
		\label{eq: unobserved prob 2}
		\end{align}
		assuming $\text{var}(L^0|\bfC) = \text{var}(L^1|\bfC)$.  Comparing  (\ref{eq: unobserved prob 1}) to (\ref{eq: unobserved prob 2}) , we obtain 
		\[\qonel = \alpha L^1 =  (\rho-1)\omega_3L^1,\]
		and together with (\ref{eq: switching prob})  we conclude that the effect of $L^1$ on switching in the control group is $\rho$ times weaker than the effect of $L^0$ on switching in the control group, conditional on the baseline covariates. The same reasoning holds if model (\ref{eq: switching prob}) uses an exponential link function instead of an expit link function.  However, the settings under which this choice of $\qone$ is correct are not limited to the settings discussed here. In particular, this choice can also be correct in other settings, where $\bfL^0$ and  $\bfL^1$  may not be normally distributed, or switching is not rare. 
		  This motivates us to propose the following sensitivity analysis for $\qone$: 
		\begin{enumerate}
			\item Fit a model for the probability of switching in the treatment group:
			\[\lm = P(S = 1|\bfL^1, R = 1,\bfC),\]
			e.g.  $\lm  = \text{expit}(\omega_1 + \bfomega_2'\bfC + \bfomega_3'\bfL^1)$. Parameter $\bfomega_3'$ is defined as the coefficient of $\bfL^1$ in this model. 
			\item Let model $\qone = (\rho-1)\bfomega_3'\bfL^1$ vary for a range of values $\rho$. One could for example test a range of values  $\rho \in [0.8,1]$ since we expect a high correlation between $\bfL^1$ and $\bfL^0$ conditional on the baseline covariates. 
			\item Repeat estimation of $\mu_1$ for each choice of $\qone$.  
		\end{enumerate}
	
\section{Implementation in R}
\label{Appendix: implementation in R}
In this section, we show how data can be simulated  and how the IPW estimator, proposed in section \ref{Section: IPW}, can be implemented.  This is done in the software package R, Version 4.0.0. 

\subsection*{Data simulation} 

We choose treatment $R$ to be randomly assigned, with equal probability. Next, a baseline covariate $C$ is generated, following a standard normal distribution. Afterwards, the counterfactual severity of disease under treatment is drawn for all patients, using a normal distribution: 
\begin{equation*}
L^1 |C \sim N(\delta_1 + \delta_2C, \sigma^2_L). 
\end{equation*}
Next, $S^1$ and $Y^1$ are generated for all patients. However, we will only use these values for the patients in the treatment group. 
\begin{align*}
S^1 | L^1, R=1,C &\sim \Bin(\expit(\omega_1+\omega_2C+\omega_3L^1)) \\
Y^1| S^1, L^1,R = 1, C &\sim N(\alpha_1 + \alpha_2S^1 + \alpha_3L^1+\alpha_4C, \sigma^2_Y). 
\end{align*}
The decision to switch a patient in the control group, conditional on $L^1$ can be generated as follows: 
\begin{equation*}
S^0 | L^1, R=0,C \sim \Bin(\expit(\lambda_1+\lambda_2C+\rho\omega_3L^1)), 
\end{equation*}
with $\rho$  the correlation between $L^1$ and $L^0$ conditional on the baseline covariates (see appendix \ref{Appendix: sensitivity analysis}).
Next, the outcome for the patients in the control group can be drawn: 
\begin{equation*}
Y^0| S^0, L^1,R = 0, C \sim N( \alpha_1 + \alpha_2S^0 + \alpha_3L^1+\alpha_4C + \alpha_5, \sigma^2_Y).
\end{equation*}
Finally, a dataset with the observed data for the patients in the control and treatment group is constructed. The parameters of scenario 1, as shown in table \ref{table parameter values} in appendix \ref{Appendix: Simulation settings}, are used. 
\begin{verbatim}
delta1<- -0.5
delta2<- 0.1
sd.l<- 0.3
omega1<- -7
omega2<- -0.01
omega3<- -7
alpha1<- 0
alpha2<- 0.5
alpha3<- 2
alpha4<- 0.1
alpha5<- -0.5
lambda1<- -5
lambda2<- -0.02
sd.y<- 0.3
rho<- 0.9
\end{verbatim}
We set a seed to make the code reproducible and choose total sample size $n = 1000$.
\begin{verbatim}
install.packages("clusterPower")
library("clusterPower")
set.seed(123)
n<-1000 
R<-rbinom(n,1,0.5) 
C<-rnorm(n,0,1) 
L.one<-rnorm(n,delta1+delta2*C,sd.l) 
#treatment group
S.one<-rbinom(n,1,expit(omega1+omega2*C+omega3*L.one)) 
Y.one<-rnorm(n,alpha1+alpha2*S.one+alpha3*L.one+alpha4*C,sd.y) 
#control group
S.zero<-rbinom(n,1,expit(lambda1+lambda2*C+rho*omega3*L.one))
Y.zero<-rnorm(n,alpha1+alpha2*S.zero+alpha3*L.one+alpha4*C+alpha5,sd.y)
#observed data
L<-ifelse(R==1,yes=L.one,no=NA)
S<-R*S.one+(1-R)*S.zero
Y<-R*Y.one+(1-R)*Y.zero
data<-data.frame(R,C,L,S,Y)
\end{verbatim}	
The following display shows the data structure:
\begin{verbatim}
head(data)
R           C          L S             Y
1 0 -0.60189285         NA 0 -1.9228484239
2 1 -0.99369859 -0.6915470 1 -1.0807910570
3 0  1.02678506         NA 1 -0.8404511436
4 1  0.75106130 -0.2367732 0 -0.6076257387
5 1 -1.50916654 -0.3148101 0 -0.0009898745
6 0 -0.09514745         NA 0  0.1251228824
\end{verbatim}
For the patients in the control group ($R = 0$), $L$ is missing. This is not a problem, since the observed $L$ in the control group  is not used to estimate $\mu$. However, to make the rest of the implementation easier, we will replace these values by 0. 
\begin{verbatim}
L<-ifelse(R==1,yes=L,no=0)
data$L<-L
head(data)
R           C          L S             Y
1 0 -0.60189285  0.0000000 0 -1.9228484239
2 1 -0.99369859 -0.6915470 1 -1.0807910570
3 0  1.02678506  0.0000000 1 -0.8404511436
4 1  0.75106130 -0.2367732 0 -0.6076257387
5 1 -1.50916654 -0.3148101 0 -0.0009898745
6 0 -0.09514745  0.0000000 0  0.1251228824
\end{verbatim}
The treatment policy estimand or intention-to-treat effect can easily be calculated: 
\begin{verbatim}
mean(subset(data,R==1)$Y)-mean(subset(data,R==0)$Y)
0.4001212
\end{verbatim}
In the next section, we illustrate how the novel estimand  can be estimated using the IPW estimator proposed in section \ref{Section: IPW}. 
\subsection*{IPW estimator}
Step 1: fit model $l(L,C,\bfomega)$ for the probability of switching in the treatment group $P(S = 1|L,R = 1,C)$
and save the predictions. 
\begin{verbatim}
fit.l<-glm(S~L+C,data=subset(data,R==1),family = binomial(link="logit")) 
prob.l<-predict(fit.l,newdata=data.frame(C=data$C,L=data$L),type="response")
omega1.estimate<-coef(fit.l)["(Intercept)"]
omega2.estimate<-coef(fit.l)["C"]
omega3.estimate<-coef(fit.l)["L"]
\end{verbatim}
The coefficients in this model are saved to use in the next step. 

\noindent Step 2: 
Calculate sample mean of $R$:  $\hat{\pi} = n^{-1}\sum_{i = 1}^n R_i$ for all patients across both treatment groups.
\begin{verbatim}
prob.pi<-dim(subset(data,R==1))[1]/dim(data)[1]
\end{verbatim}
Estimate parameter $\bflambda = (\lambda_1, \lambda_2)'$ by solving estimating equations. 
Therefore, we define a function that takes a value for $\lambda_1$  and $\lambda_2$ as input 
and returns the obtained values for 
\begin{eqnarray*}
	&&\mathbf{0}	= \sum_{i = 1}^n \begin{pmatrix}
		1 \\ C_i
	\end{pmatrix}  \left( \frac{(1-R_i)(1-S_i)}{1-\hat{\pi}} \right. \\
	&& \left. -\frac{1}{\hat{\pi}}  \frac{R_i(1-S_i)}
	{\text{expit}(\hat{\omega}_1 + \hat{\omega}_2C_i + \hat{\omega}_3L_i)\left\{\exp\left(\lambda_1 - \hat{\omega}_1+ (\lambda_2 -\hat{\omega}_2)C_i  + (\rho-1)\hat{\omega}_3L_i \right)-1 \right\} +1 }\right) ,
\end{eqnarray*} 
corresponding to these  $\lambda_1$  and $\lambda_2$. 
\begin{verbatim}
IPW.lambda<-function(x){
lambda1<-as.numeric(x[1])
lambda2<-as.numeric(x[2])
q0<-lambda1-omega1.estimate+(lambda2-omega2.estimate)*C
q1<-(rho-1)*omega3.estimate*L
main<-(1-R)*(1-S)/(1-prob.pi)-(R/prob.pi)*(1-S)/(prob.l*(exp(q0+q1)-1)+1)
return(c(sum(main),sum(C*main)))
}
\end{verbatim}
Now $\lambda_1$  and $\lambda_2$ can be estimated by solving the estimating equations using the ${\tt nleqslv}$ function. We choose $\lambda_1 =0$  and $\lambda_2=0$ as start values for the algorithm. 
\begin{verbatim}
install.packages("nleqslv")
library("nleqslv")
lambda<-nleqslv(fn=IPW.lambda, x=c(0,0))$x
\end{verbatim}
Step 3: estimate the weights $W(S,L,C)$.
\begin{verbatim}
q0<-lambda[1]-omega1.estimate+(lambda[2]-omega2.estimate)*C
q1<-(rho-1)*omega3.estimate*L
W<-S*exp(q0+q1)/(prob.l*(exp(q0+q1)-1)+1)+(1-S)/(prob.l*(exp(q0+q1)-1)+1)
\end{verbatim}
Step 4: estimate $\mu_1$.
\begin{verbatim}
mu1<-mean(Y*R*W)/mean(R*W)
mu1
-0.8871583
\end{verbatim}
Step 6: estimate $\mu_0$.
\begin{verbatim}
mu0<-mean((1-R)*Y)/mean(1-R)
mu0
-1.354372
\end{verbatim}
Finally, $\mu$ can be estimated. 
\begin{verbatim}
mu<-mu1-mu0
mu
0.4672135
\end{verbatim}

\section{Simulations}
	\label{Appendix: Simulations}
	
	\subsection{Scenarios}
	\label{Appendix: Simulation settings}

	\begin{table}[H]
		\centering
		\makebox[\textwidth][c]{
			\begin{tabular}{c c    c c    } 
				\toprule
				Parameter & Scenario 1 & Scenario 2 & Scenario 3 \\
				\midrule 
				$\delta_1$ &  -0.5 &  -0.5 &  -0.5\\
				$\delta_2$ & 0.1 & 0.1 & 0.2\\
				$\sigma_L$ & 0.3 & 0.3&0.3\\
				$\omega_1$ & -7 & -9&-7\\
				$\omega_2$ & -0.01 & -0.01& -0.01\\
				$\omega_3$ & -7 & -12 &-11\\				
				$\alpha_1$ & 0 & 0& 0\\
				$\alpha_2$ & 0.5 & 0.5 & 0.7\\
				$\alpha_3$ & 2 &2 &2\\
				$\alpha_4$ & 0.1 & 0.1 &0.1\\
				$\alpha_5$ & -0.5 & -0.4 &-0.7\\
				$\sigma_Y$ & 0.3 & 0.3 & 0.3\\
			    $\lambda_1$ & -5 &-5 & -2\\
			    $\lambda_2$ & -0.02 & -0.02&-0.02\\
			    \midrule
			    \multicolumn{4}{c} {Balanced estimand ($\rho = 0.9$)} \\				
			    $\mu_1$ & -0.879 & -0.728 & -0.462\\
			    $\mu_0$ & -1.379 & -1.129 & -1.161\\
			    $\mu$  &  0.500 & 0.401 &0.699\\
			    \midrule
			    \multicolumn{4}{c}{Treatment policy estimand} \\
			    $E(Y^1)$ & -0.946 & -0.881 & -0.744 \\
			    $E(Y^0)$ & -1.379 & -1.129 & -1.161 \\
			    $E(Y^1-Y^0)$ &  0.433 &  0.248 & 0.417 \\				
				\bottomrule
			\end{tabular}
		}
		\caption{The parameter values for the different scenarios, together with the corresponding balanced  and treatment policy estimand.} 
		\label{table parameter values}
	\end{table}

\begin{figure}[H]
	\captionsetup[subfigure]{labelformat=empty}
	\centering
	\subfloat[][]{\includegraphics[width=0.45\textwidth]{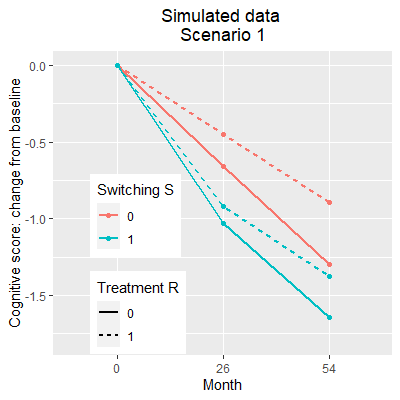}}
	\subfloat[][]{\includegraphics[width=0.45\textwidth]{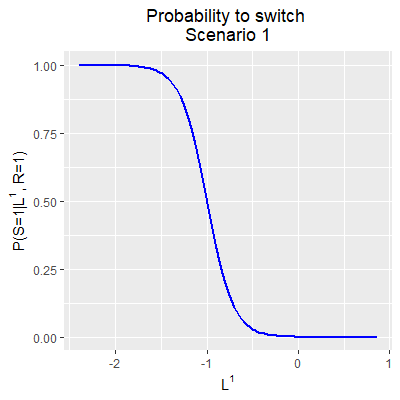}} \\
	\vspace{-30pt}
	\subfloat[][]{\includegraphics[width=0.45\textwidth]{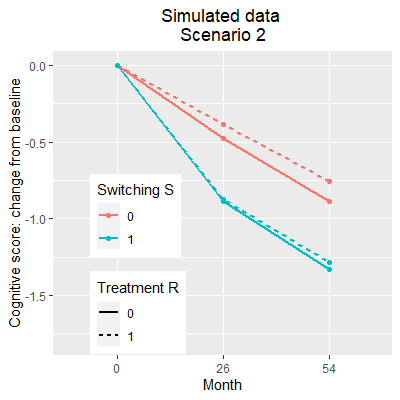}}
	\hspace{10pt}
	\subfloat[][]{\includegraphics[width=0.45\textwidth]{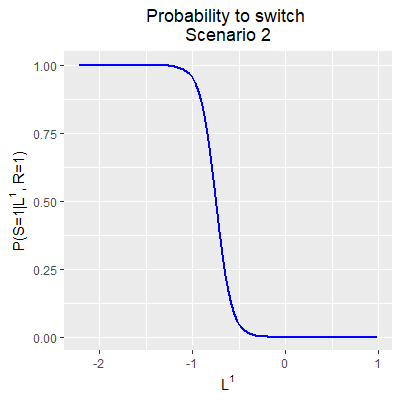}} \\
	\vspace{-30pt}
	\subfloat[][]{\includegraphics[width=0.45\textwidth]{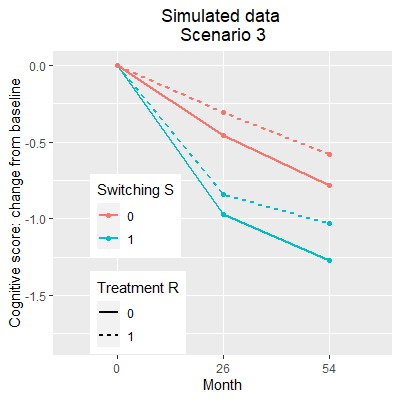}}
	\hspace{10pt}
	\subfloat[][]{\includegraphics[width=0.45\textwidth]{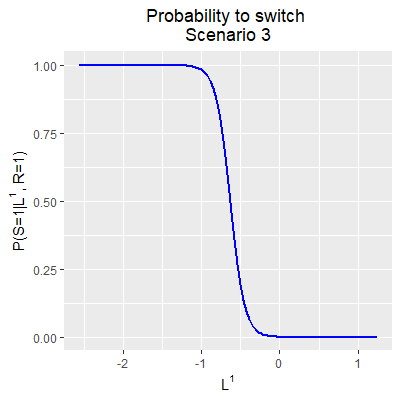}}
	\caption{Illustration of the three scenarios. At the left pictures, the mean change from baseline in the cognitive score is shown. The change in score at month 26 is the health status $L$, while the change in score at month 54 is the outcome $Y$. At the right pictures, the probability to switch in function of $L^1$ is displayed. 
	 } 
	\label{fig: scenarios}
\end{figure}

\begin{figure}[H]
	\captionsetup[subfigure]{labelformat=empty}
	\centering
	\subfloat[][]{\includegraphics[width=0.45\textwidth]{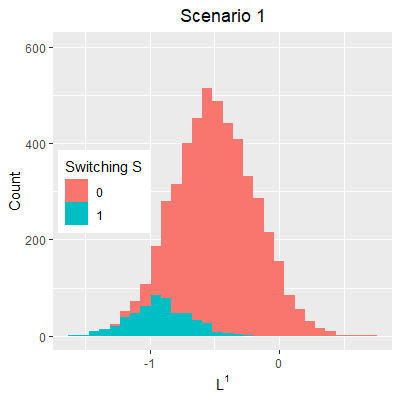}}
	\vspace{-30pt}
	\subfloat[][]{\includegraphics[width=0.45\textwidth]{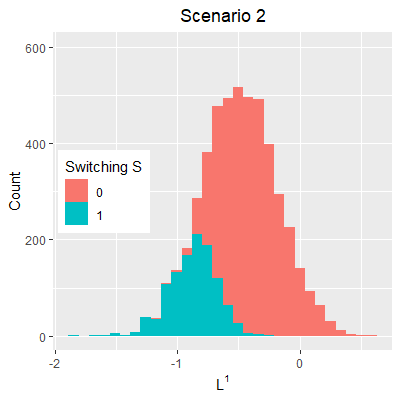}}
	\\
	\subfloat[][]{\includegraphics[width=0.45\textwidth]{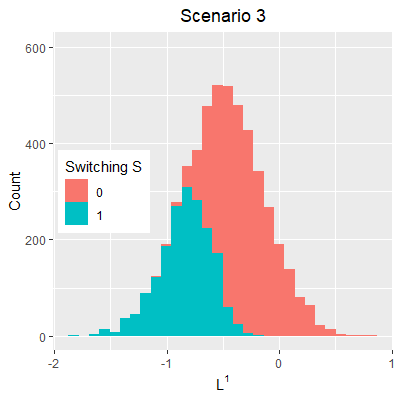}}
	\caption{Illustration of the three scenarios. The histograms of $L^1$ are shown. } 
	\label{fig: scenarios2}
\end{figure}

\subsection{Results}
\label{Appendix: Simulation results}
	\begin{figure}[H]
		\centering
		\includegraphics[width=1\textwidth]{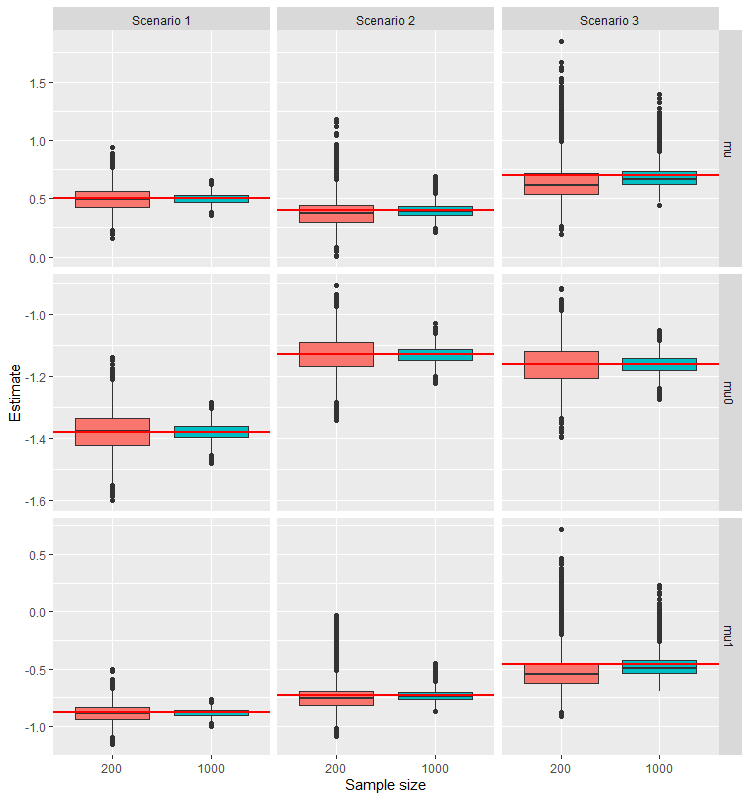}
		\caption{Boxplots showing the results of the simulations investigating the finite sample size performance of the IPW estimator for $\mu_0$, $\mu_1$ and $\mu = \mu_1 - \mu_0$. The red horizontal lines indicate the true parameter values. }
		\label{fig: results simulations}
	\end{figure}

\begin{sidewaystable}
	\centering
	\renewcommand{\arraystretch}{1}
	\begin{tabular}{lllll@{\hspace{0.4cm}}ll@{\hspace{0.4cm}}ll@{\hspace{0.4cm}}lcc}
		\toprule 
		&& \multicolumn{7}{c}{Balanced estimand} &\multicolumn{3}{c}{ Treatment policy} \\
		\cmidrule(l{2pt}r{12pt}){3-9}	 \cmidrule(l{2pt}r{12pt}){10-12}	
		  &  &  & \multicolumn{4}{c}{$\rho$ misspecified}   & \multicolumn{2}{c}{$\rho$ correct }  \\ 
		   \cmidrule(l{2pt}r{12pt}){4-7} \cmidrule(l{2pt}r{2pt}){8-9}
		  & Sample &  & \multicolumn{2}{c}{$\rho = 0.8$} &\multicolumn{2}{c}{$\rho = 1$} & \multicolumn{2}{c}{$\rho = 0.9$} \\
		Scenario &size & Parameter& Bias & SE & Bias & SE  & Bias & SE & Parameter & Bias & SE\\
		\midrule
		\textbf{Scenario 1} & 200 & $\mu$ & -0.004 & 0.102 & -0.002 & 0.101 & -0.003 & 0.101 & $E(Y^{1}-Y^0)$  &  0 & 0.096\\ 
		\multirow{5}{*}{\makecell[l]{ Large treatment effect \\ Limited  switching effect \\ Switchers: \\11\%  in $R=1$ \\24\%  in $R=0$ }}&& $\mu_1$   & -0.004 & 0.081 & -0.002 & 0.079 & -0.003 & 0.080  & $E(Y^{1})$ & 0.001 & 0.069\\ 
		&& $\mu_0$ & 0.001 & 0.064 & 0.001 & 0.064 & 0.001 & 0.064  & $E(Y^{0})$ & 0.001 & 0.067\\
		\cmidrule(lr){2-12} 
		& 1000 & $\mu$ & 0 & 0.045 & 0&0.044 & 0 & 0.044 & $E(Y^{1}-Y^0)$ & 0&  0.043 \\ 
		&& $\mu_1$ & 0 & 0.035 & 0& 0.034 & 0&0.034 &  $E(Y^{1})$ & 0 & 0.030 \\ 
		&& $\mu_0$  & 0 & 0.028 & 0 & 0.028 & 0& 0.028 &$E(Y^{0})$ & 0 & 0.029\\ 
		\cmidrule(lr){1-12} 
		\textbf{Scenario 2} & 200 & $\mu$  & -0.024 & 0.125 & -0.019 & 0.120 &-0.021&0.122 &$E(Y^{1}-Y^0)$ & -0.001 & 0.089 \\
		\multirow{5}{*}{\makecell[l]{ Small treatment effect \\ Limited  switching effect \\ Switchers: \\ 24\%  in $R=1$ \\ 54\%  in $R=0$ }}&& $\mu_1$ & -0.024 & 0.110&-0.018&0.104&-0.020&0.106 & $E(Y^{1})$ & -0.001 & 0.064 \\ 
		&& $\mu_0$ & 0.001 & 0.059 &0.001&0.059&0.001&0.059 & $E(Y^{0})$ & 0.001 &0.062 \\ 
		\cmidrule(lr){2-12}  
		& 1000 & $\mu$ & -0.005&0.062 &-0.003&0.055 &-0.004&0.058 &$E(Y^{1}-Y^0)$ & -0.001 & 
		0.040  \\ 
		&& $\mu_1$ &-0.004&0.056& -0.003&0.048&-0.003&0.051 & $E(Y^{1})$ & 0 &	0.028 \\ 
		&& $\mu_0$  & 0&0.026&0&0.026&0&0.026 & $E(Y^{0})$ & 0 & 
		0.027\\ 
		\cmidrule(lr){1-12} 
		\textbf{Scenario 3}  & 200 & $\mu$ & -0.067 & 0.168 & -0.059 & 0.162 & -0.062 & 0.164 &$E(Y^{1}-Y^0)$ & 0 & 
		0.098\\
		\multirow{5}{*}{\makecell[l]{ Large treatment effect \\ Large  switching effect \\ Switchers: \\36\%  in $R=1$ \\76\%  in $R=0$ }}&& $\mu_1$  & -0.066&0.160 & -0.059 & 0.153 & -0.062 & 0.156 & $E(Y^{1})$ & 0 & 0.067\\ 
		&& $\mu_0$ & 0.001&0.065 & 0.001 & 0.065 & 0.001 & 0.065 & $E(Y^{0})$ & 0.001 &	0.071 \\ 
		\cmidrule(lr){2-12} 
		& 1000 & $\mu$  & -0.014&0.113 & -0.010 & 0.100 & -0.012 & 0.106&$E(Y^{1}-Y^0)$ & 0 &0.044  \\ 
		& & $\mu_1$ & -0.014&0.110  & -0.010 & 0.096 & -0.011 & 0.102 & $E(Y^{1})$ & 0 & 0.029 \\ 
		&& $\mu_0$ & 0& 0.029 & 0& 0.029 & 0 & 0.029 &$E(Y^{0})$ & 0 & 
		0.032 \\ 
		\hline
	\end{tabular}
\caption{Results of the simulations investigating the finite sample size performance of the treatment policy estimand and the IPW estimator proposed in section \ref{Section: IPW}, for three different values of $\rho$. }
	\label{table: Results simulations different rho}
\end{sidewaystable}

\begin{table}[H]
	\centering
	\renewcommand{\arraystretch}{1}
	\begin{tabular}{llllll}
		\cmidrule(lr){1-6} 
		Scenario  & Sample  & Parameter & Bias & SE & Weights  \\ 
		& size  &  \\ 
		\cmidrule(lr){1-6} 
		\textbf{Scenario 1} & 200 & $\mu$  & -0.008 & 0.098 & \\ 
		\multirow{5}{*}{\makecell[l]{ Large treatment effect \\ Limited  switching effect \\ Switchers: \\11\%  in $R=1$ \\24\%  in $R=0$ }}&& $\mu_1$  & -0.007 & 0.076 & 0.437; 1.642  \\ 
		&& $\mu_0$  & 0.001 & 0.064 & \\
		\cmidrule(lr){2-6} 
		& 1000 & $\mu$  & -0.006 & 0.043 & \\ 
		&& $\mu_1$ &  -0.006 & 0.033 & 0.508; 1.895  \\ 
		&& $\mu_0$  & 0 & 0.028 & \\ 
		\cmidrule(lr){1-6} 
		\textbf{Scenario 2} & 200 & $\mu$  & -0.042 & 0.099 &\\
		\multirow{5}{*}{\makecell[l]{ Small treatment effect \\ Limited  switching effect \\ Switchers: \\ 24\%  in $R=1$ \\ 54\%  in $R=0$ }}&& $\mu_1$  &-0.042 & 0.081 & 0.054; 2.132   \\ 
		&& $\mu_0$  & 0.001 & 0.059 & \\ 
		\cmidrule(lr){2-6}  
		& 1000 & $\mu$  & -0.035 & 0.044 & \\ 
		&& $\mu_1$ & -0.034 & 0.035 & 0.088; 2.240  \\ 
		&& $\mu_0$  & 0 & 0.026 & \\ 
		\cmidrule(lr){1-6} 
		\textbf{Scenario 3}  & 200 & $\mu$ & -0.123 & 0.114 & \\
		\multirow{5}{*}{\makecell[l]{ Large treatment effect \\ Large  switching effect \\ Switchers: \\36\%  in $R=1$ \\76\%  in $R=0$ }}&& $\mu_1$  & -0.122 & 0.103 & 0.010; 2.398 \\ 
		&& $\mu_0$ & 0.001 & 0.065& \\ 
		\cmidrule(lr){2-6} 
		& 1000 & $\mu$  & -0.105& 0.052 &  \\ 
		& & $\mu_1$ & -0.105 & 0.046 &0.023; 2.424   \\ 
		&& $\mu_0$ & 0 & 0.029 &\\ 
		\hline
	\end{tabular}
	\caption{Results of the simulations investigating the finite sample size performance of the IPW estimator proposed in section \ref{Section: IPW}, with truncated weights $W_i$ at the 1\% and 99\% percentiles.  The column `weights' shows the 5\% and 95\% percentiles of the weights 
		$W_i/n^{-1} \sum_{j = 1}^n R_jW_j  $	among the patients in the experimental treatment arm.  }
	\label{table: Results simulations truncated weights}
\end{table}

\section{Data Analysis}
\subsection{Estimands framework}
\label{Appendix: Estimands framework data analysis}

    The balanced, treatment policy and hypothetical estimand used in the data analysis (section \ref{Section: data analysis} of the main paper) can be defined according to the guidelines of the ICH E9(R1) addendum. These estimands only differ in the way in which the intercurrent event `switching to rescue medication' is handled:

	\begin{itemize}
	\item \textbf{Treatment}: canagliflozin 100 mg or placebo, as defined by the study protocol. 
	\item \textbf{Population}: the entire study population, as defined by the inclusion-exclusion criteria of the study. 
	\item \textbf{Variable}: change from baseline in HbA1c (\%) at week 26. 
	\item \textbf{Intercurrent events}:
	\begin{itemize}
		\item \emph{Study discontinuation}
		
		The hypothetical scenario is envisaged where patients would not discontinue.
		
		\item \emph{Switching to rescue medication}
		
		\begin{itemize}
			\item Balanced estimand:
			
			The hypothetical scenario is envisaged where patients on the placebo arm had been switched to rescue medication if and only if they would have been switched when randomised to canagliflozin 100 mg. 
			
			\item Treatment policy estimand:
			
			All observed values of the variable are of interest, regardless of whether or not the patient had initiated rescue medication. 
			
			\item Hypothetical estimand: 
			
			The hypothetical scenario is envisaged where patients would not switch to rescue medication. 
		\end{itemize}

	\end{itemize} 
	\item \textbf{Population-level summary}: difference in means of the variable. 
\end{itemize}

\subsection{Materials and methods}
\label{Appendix: Materials and methods}

\begin{table}[H]
	\centering
	\renewcommand{\arraystretch}{1}
	\begin{tabular}{lllll}
		\toprule
	 & Placebo  & CANA 100 mg  & CANA 300 mg & Total \\ 	
	Characteristic &  (n = 192) &  (n = 195) & (n = 197) &  (n = 584)\\
	\cmidrule{1-5}
	Sex, n (\%) \\
	\hspace*{2pt} Male & 88 (45.8) & 81 (41.5) & 89 (45.2) & 258 (44.2) \\
	\hspace*{2pt} Female & 104 (54.2) & 114 (58.5) & 108 (54.8) & 326 (55.8) \\
	Race, n (\%) \\ 
		\hspace*{2pt} White & 134 (69.8) & 124 (63.6) & 137 (69.5) & 395 (67.6) \\
		\hspace*{2pt} Black or African & 9 (4.7) & 18 (9.2) & 14 (7.1) & 41 (7.0) \\
		\hspace*{2pt} Asian & 29 (15.1) & 27 (13.8) & 29 (14.7) & 85 (14.6) \\
		\hspace*{2pt} Other & 20 (10.4) & 26 (13.3) & 17 (8.6) & 63 (10.8) \\
	Age (years), Mean (sd) & 55.7 (10.9) & 55.1 (10.8) & 55.3 (10.2) & 55.4 (10.6) \\
	HbA1c (\%), Mean (sd) & 8.0 (1.0) & 8.1 (1.0) & 8.0 (1.0) & 8.0 (1.0)\\
	FPG (mmol/l), Mean (sd) & 9.3 (2.1) & 9.6 (2.4) & 9.6 (2.4) & 9.5 (2.3) \\
	Body weight (kg), Mean (sd) &  87.6 (19.5) & 85.8 (21.4) & 86.9 (20.5) & 86.8 (20.4) \\
	BMI (kg/$\text{m}^2$),  Mean (sd) & 31.8 (6.2) & 31.3 (6.6) & 31.7 (6.0) & 31.6 (6.2)\\
	Duration of diabetes & 25 (13.0) & 23 (11.8) & 20 (10.2) & 68 (11.6) \\ 
	$\geq$  10 years,  n (\%) \\
	On AHA at screening, n (\%) & 92 (47.9) & 94 (48.2) & 95 (48.2) & 281 (48.1)  \\
	Participating  & 76 (39.6) & 79 (40.5) & 80(40.6) & 235 (40.2) \\
	in FS-MMTT, n (\%) \\
	 GFR (ml/min/1.73m$^2$),& 86.0 (21.5) & 88.5 (20.2) & 86.6 (19.1) & 87.1 (20.3)\\
	 Mean (sd) \\
	Smoker, n (\%) & 22 (11.5) &34 (17.4) & 21 (10.7) & 77 (13.2)\\
	Systolic BP $\geq$ 140 mmHg, & 46 (24.0) & 52 (26.7) & 51 (25.9) &149 (25.5)\\
	 n (\%) \\
	SHDL (mmol/l), Mean (sd) & 1.1 (0.3)  & 1.2 (0.3) & 1.2 (0.3) & 1.2 (0.3)\\
	STRIG (mmol/l), Mean (sd) & 2.3 (1.3) & 2.0 (1.3) & 2.0 (1.1) & 2.1 (1.2) \\
	PPG (mmol/l), Mean (sd) & 13.8 (3.9) & 14.2 (4.4) & 14.2 (4.1) & 14.1 (4.2) \\
	\bottomrule
 	\end{tabular}
	\caption{Baseline and disease characteristics of patients  in the canagliflozin study. \\
	AHA, antihyperglycaemic agent; BMI, body mass index; BP, blood pressure; CANA, canagliflozin; FPG, fasting plasma glucose; FS-MMTT, frequently-sampled mixed-meal tolerance
	test;  GFR, glomerular filtration
	rate; HbA1c, haemoglobin A1c; PPG, plasma glucose - MMTT at 2 hour;   SHDL, serum HDL cholesterol; STRIG, serum triglycerides.}
	\label{table: Baseline characteristics}
\end{table}

\begin{table}[H]
	\centering
	\renewcommand{\arraystretch}{1}
	\begin{tabular}{llll}
		\toprule
		& Placebo  & CANA 100 mg  & Total \\ 	
		Characteristic &  (n = 192) &  (n = 195)  &  (n = 387)\\
	   \cmidrule{1-4}
	   FPG \\
	   \hspace*{2pt} Baseline  &  0 &0&0\\
	   \hspace*{2pt} Week 6 &  9 (4.6) & 8 (4.1) & 17 (4.4) \\
	   \hspace*{2pt} Week 12 & 15 (7.8) & 16 (8.2) & 31 (8.0) \\
	   \hspace*{2pt} Week 18 &  23 (12.0) & 20 (10.3) & 43 (11.1)\\
	   \hspace*{2pt} Week 26 & 60 (31.3) & 35 (18.0) & 95 (24.5) \\
		HbA1c \\
		\hspace*{2pt} Baseline  &  0 &0&0\\
		\hspace*{2pt} Week 6  &  3 (1.6) & 7 (3.6) & 10 (2.6)\\
		\hspace*{2pt} Week 12 & 12 (6.3) & 13 (6.7) & 25 (6.5)  \\
		\hspace*{2pt} Week 18&  19 (9.9) & 17 (8.7) & 36 (9.3) \\
		\hspace*{2pt} Week 26 & 29 (15.1) & 25 (12.8) & 54 (14.0)\\
		PPG \\
		\hspace*{2pt} Baseline  &  10 (5.2) & 6 (3.1) & 16 (4.1)\\
		\hspace*{2pt} Week 26 & 72 (37.5) & 39 (20)  & 111 (28.7)\\
		SHDL \\
		\hspace*{2pt} Baseline  &  0 &0&0\\
		\hspace*{2pt} Week 26 & 30 (15.6) & 26 (13.3) & 56 (14.5)\\
	    STRIG \\
		\hspace*{2pt} Baseline  &  0 &0&0\\
		\hspace*{2pt} Week 26 & 30 (15.6) & 25 (12.8) & 55 (14.2) \\
		\bottomrule
	\end{tabular}
	\caption{Number (\%) of missing values in the longitudinal variables.  \\
		 CANA, canagliflozin; FPG, fasting plasma glucose; HbA1c, haemoglobin A1c; PPG, plasma glucose - MMTT at 2 hour;   SHDL, serum HDL cholesterol; STRIG, serum triglycerides.}
	\label{table: Missing data}
\end{table}
	

\begin{figure}[H]
	\captionsetup[subfigure]{labelformat=empty}
	\centering
	\subfloat[][]{\includegraphics[width=0.5\textwidth]{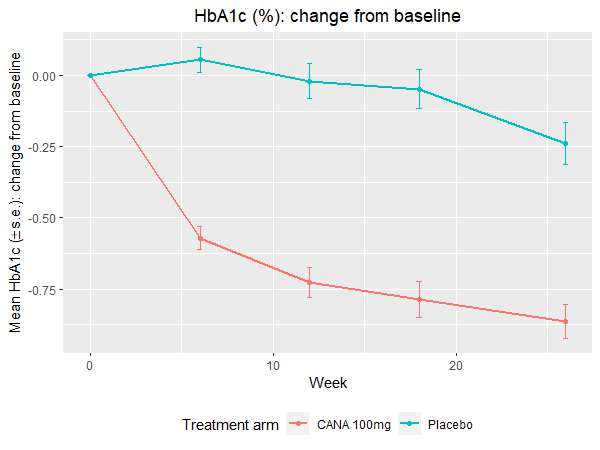}}
	\subfloat[][]{\includegraphics[width=0.5\textwidth]{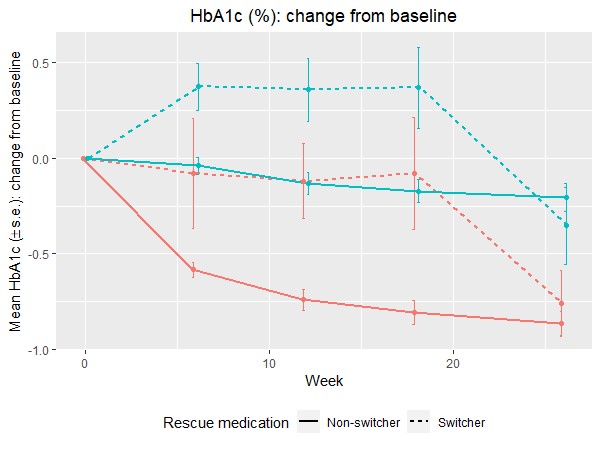}}
	\caption{Mean change in HbA1c over time based on the first imputed dataset. Plots for the other imputed datasets are very similar.} 
	\label{fig: results mean HbA1C}
\end{figure}

\begin{figure}[H]
	\centering
	\subfloat[Complete data]{\includegraphics[width=0.5\textwidth]{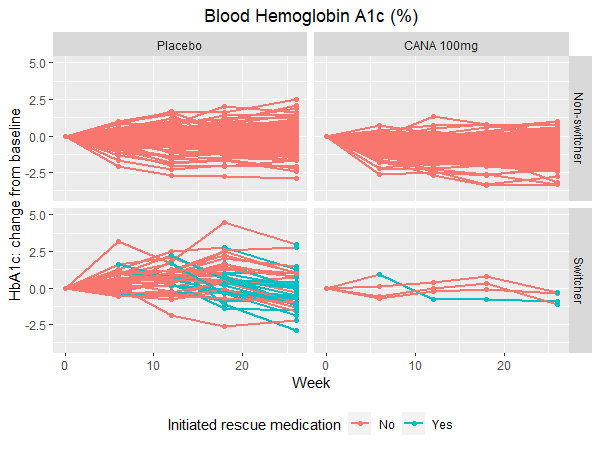}\label{fig: b1}}
	\subfloat[First imputed dataset]{\includegraphics[width=0.5\textwidth]{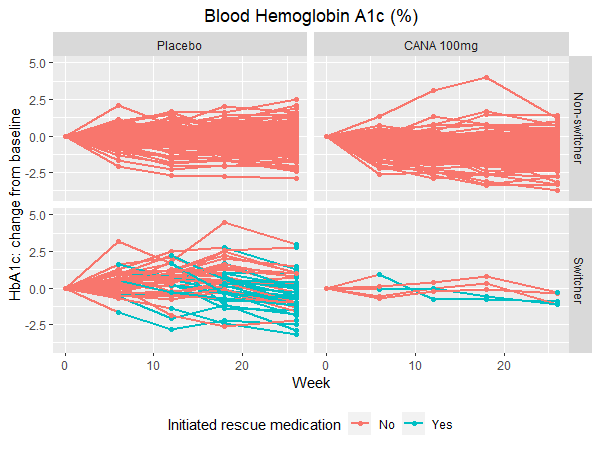}\label{fig: b2}}
	\caption{Change in HbA1c over time for  (a) the patients without missing HbA1c values (b) the first imputed dataset. The line turns green from the first study visit after initiation of rescue medication. Plots for the other imputed datasets are very similar.} 
	\label{fig: results time HbA1C}
\end{figure}

\begin{figure}[H]
	\centering
	\subfloat[Complete data]{\includegraphics[width=0.5\textwidth]{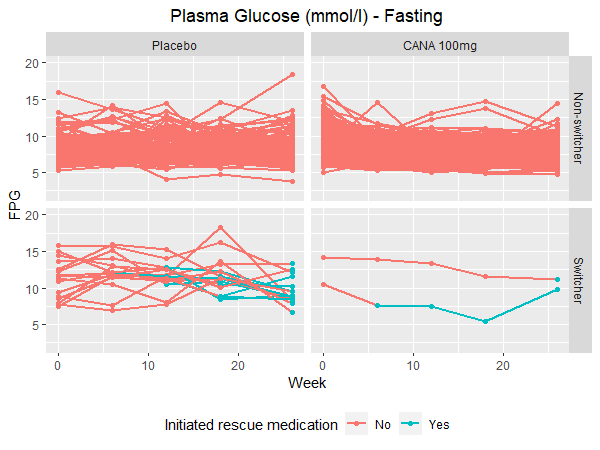}\label{fig: b1}}
	\subfloat[First imputed dataset]{\includegraphics[width=0.5\textwidth]{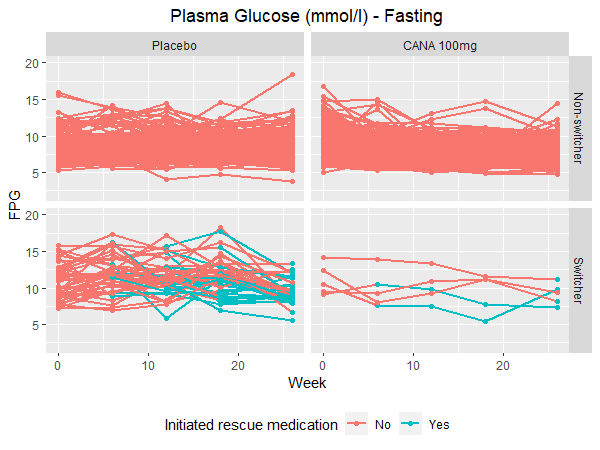}\label{fig: b2}}
	\caption{Change in FPG over time for  (a) the patients without missing FPG values (b) the first imputed dataset. The line turns green from the first study visit after initiation of rescue medication. Plots for the other imputed datasets are very similar.  }
	\label{fig: results time FPG}
\end{figure}

\subsection{Treatment policy estimand}
\label{Appendix: Treatment policy estimand}

In the treatment policy or intention-to-treat estimand regarding the intercurrent event `switching to rescue medication', the initiation of rescue medication is considered irrelevant in defining the treatment effect of interest:
$E(Y^{1} - Y^{0})$.

Assuming consistency and no unmeasured confounders between treatment and outcome ($Y^{r} \independent R |\bfC $ for $ r \in \{0,1\}$), as is guaranteed by randomised assignment, the treatment policy estimand can be expressed as
\begin{align*}
	E(Y^{1} - Y^{0}) = E\left(Y \frac{ R}{P(R=1|\bfC)}\right) -E\left(Y \frac{ 1-R}{P(R=0|\bfC)}\right).
\end{align*}
Therefore, we propose the following IPW estimator for the treatment policy  estimand: 
\begin{align*}
\sum_{i = 1}^n Y_i \frac{R_i}{\hat{\pi}(\bfC_i)}  / \sum_{i = 1}^n  \frac{R_i}{\hat{\pi}(\bfC_i)}  
-\sum_{i = 1}^n Y_i \frac{1-R_i}{1-\hat{\pi}(\bfC_i)}  / \sum_{i = 1}^n  \frac{1-R_i}{1-\hat{\pi}(\bfC_i)}.  \\
\end{align*}
In the data-analysis, the same 50 imputed datasets were used as for the estimation of the balanced estimand. For  each of these datasets, the treatment policy effect was estimated and standard errors are obtained from  1000 nonparametric stratified bootstrap replications. The results are combined using Rubin's rules.   The 95\% confidence intervals are estimated as percentile intervals from the pooled sample of 50$\times$1000 estimates of the treatment policy effect. 

\subsection{Hypothetical estimand}
\label{Appendix: Hypothetical estimand}
The hypothetical estimand regarding the intercurrent event `switching to rescue medication' corresponds to the hypothetical scenario where patients would not switch to rescue medication: $E(Y^{10} - Y^{00})$.

Identification of this hypothetical estimand relies on  consistency and  assumptions
\begin{align*}
Y^{r0} &\independent R | \bfL^r, \bfC \\
\bfL^r &\independent R |\bfC \\
Y^{r0} &\independent S | \bfL, R, \bfC. 
\end{align*}
The first two assumptions are guaranteed by randomisation. The last assumption states that the complier status of a patient  can be fully attributed to the baseline covariates, measured confounders and the treatment group in the sense that it has no residual dependence on the potential
outcomes $Y^{r0}$. Using these assumptions, the hypothetical estimand can be expressed as
\begin{align*}
&E(Y^{10} - Y^{00}) \\
&= \int E(Y^{10}|\bfL^1, \bfC) f(\bfL^1|\bfC)f(\bfC)d\bfL^1d\bfC 
- \int E(Y^{00}|\bfL^0, \bfC) f(\bfL^0|\bfC)f(\bfC)d\bfL^0d\bfC \\
&= E\left(Y \frac{R}{P(R=1|\bfC)} \frac{1-S}{P(S = 0|\bfL, R = 1, \bfC)}\right) -E\left(Y \frac{1-R}{P(R=0|\bfC)} \frac{1-S}{P(S = 0|\bfL, R = 0, \bfC)}\right).
\end{align*}
Therefore, we propose the following IPW estimator for the hypothetical estimand: 
\begin{align*}
&\sum_{i = 1}^n Y_i \frac{R_i}{\hat{\pi}(\bfC_i)} \frac{1-S_i}{\hat{P}(S_i = 0|\bfL_i, R_i = 1, \bfC_i)} / \sum_{i = 1}^n  \frac{R_i}{\hat{\pi}(\bfC_i)} \frac{1-S_i}{\hat{P}(S_i = 0|\bfL_i, R_i = 1, \bfC_i)} \\
&-\sum_{i = 1}^n Y_i \frac{1-R_i}{1-\hat{\pi}(\bfC_i)} \frac{1-S_i}{\hat{P}(S_i = 0|\bfL_i, R_i = 0, \bfC_i)} / \sum_{i = 1}^n  \frac{1-R_i}{1-\hat{\pi}(\bfC_i)} \frac{1-S_i}{\hat{P}(S_i = 0|\bfL_i, R_i = 0, \bfC_i)}, \\
\end{align*}
where $\hat{P}(S_i = 0|\bfL_i, R_i = 1, \bfC_i)$ and  $\hat{P}(S_i = 0|\bfL_i, R_i = 0, \bfC_i)$ are obtained by regressing the switching status on the variables $\bfL$ and $\bfC$ in the treatment or placebo group using logistic regression. The probability to switch in the placebo group, i.e. $P(S=1|\bfL,R=0,\bfC)$, is estimated using backward elimination, in the same way as for the balanced estimand (see section \ref{Section: data analysis} in the main paper). 
Since in the full model for the probability to switch in the treatment group, i.e.
$P(S=1|\bfL,R=1,\bfC)$, all predictors had a $p$-value of 99\%, we decided to perform forward elimination, starting from the model with FPG  as only predictor. This logistic model was fitted in every imputed dataset and pooled using Rubin's rules. Next, in each step 1 variable was added to the model and kept if the corresponding $p$-value in the pooled model was below 10\%. Next, all two-by-two interactions between the variables in the model were each in turn added to the model and kept if the corresponding $p$-value was below 10\%. Afterwards,  the hypothetical effect was estimated for every imputed dataset and standard errors are obtained from  1000 nonparametric stratified bootstrap replications. Stratification was performed according to whether
subjects were taking AHAs at screening and whether they
participated in the frequently-sampled mixed-meal tolerance
test (FS-MMTT). 
 The results are combined using Rubin's rules.   The 95\% confidence intervals are estimated as percentile intervals from the pooled sample of 50$\times$1000 estimates of the hypothetical effect. 

\subsection{Results}
\label{Appendix: Results}

\begin{table}[h]
	\centering
	\begin{tabular}{lc}
		\toprule
	  \multicolumn{2}{l}{Dependent variable: switching status $S$}\\ 
	 \cmidrule(ll){1-2} 
		Variable & Coefficient (Standard error) \\ 
		\cmidrule(ll){1-2} 
		Intercept & $-9.228 (2.551)^{***}$\\
		AGE & $-0.064    (0.031)^*$\\
		AHASTRAT & $-1.467    (0.631)^*$ \\
		BLFPG  & $-1.083 (0.282)^{***}$\\
		SNSBPHFL & $1.786 (0.748)^*$ \\
		BLSTRIG & $-0.433 (0.243)^*$\\
		FPG & $2.249 (0.417)^{***}$\\
		\cmidrule(ll){1-2}  
		Observations & 192 \\
		\bottomrule
	\end{tabular}
	\caption{Logistic regression results for the pooled model for  $P(S=1|\bfL,R=0,\bfC)$.   \\ Predictors: AGE, age in years; AHASTRAT, stratification factor: whether or not patient was taking antihyperglycaemic agents at screening  (reference: yes);  BLFPG,  fasting plasma glucose value at baseline (mmol/l); BLSTRIG, serum triglycerides at baseline (mmol/l); FPG, average fasting plasma glucose value  (mmol/l) (for switchers: average before switching, for non-switchers: average before week 26); SNSBPHFL, systolic blood pressure  $\geq$ 140 mmHg at screening (reference: no). \\
	Significance codes: $^*p<0.05$, $^{**}p<0.01$ , $^{***}p<0.001$ }
	\label{table: Regression model l}
\end{table}

\begin{table}[h]
	\centering
	\begin{tabular}{lccc}
		\toprule
		& \multicolumn{3}{c}{Coefficient} \\
		\cmidrule(ll){2-4} 
		 Variable & $\rho = 1$ &  $\rho = 0.9 $ & $\rho = 0.8 $ \\ 
		\cmidrule(ll){1-4} 
		 Intercept & -35.826 & -33.988 & -32.305 \\
		    AGE &  0.335 &  0.322 &  0.312 \\
		    AHASTRAT & 3.061 & 3.056 &3.074 \\
		     BLFPG &-1.908 & -1.740 & -1.578 \\
		    SNSBPHFL & -4.725 & -4.457 & -4.201 \\		   		 
		 BLSTRIG & 3.008 &2.889 & 2.790  \\
		 FPG &  2.158 &1.943  & 1.727 \\
		\bottomrule
	\end{tabular}
	\caption{Logistic regression results for $P(S=1|\bfL^0,R=1,\bfC)$ for the first imputed dataset. Differences with the parameter values for the other imputed datasets are very limited. }
	\label{table: Results Functions q0 q1}
\end{table}

\begin{figure}[h]
	\centering
	\includegraphics[width=0.7\textwidth]{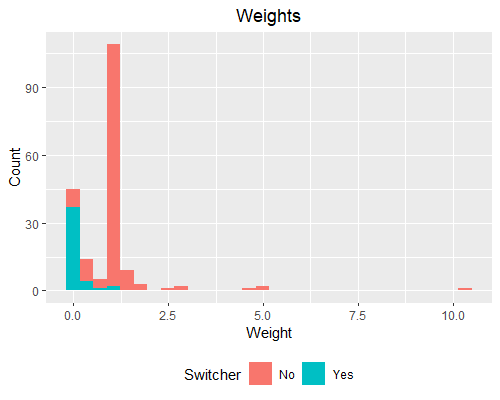}
	\caption{Histogram of the weights $W(S,\bfL,\bfC)$ used to estimate the balanced estimand ($\rho = 0.9$). Weights are shown for the patients in the placebo arm of the study using the first imputed dataset. Differences with the plots for the other imputed datasets and other values for the sensitivity parameter $\rho$ are limited.  }
	\label{fig: results weights}
\end{figure}

\begin{table}[h]
	\centering
	\begin{tabular}{lc}
		\toprule
		\multicolumn{2}{l}{Dependent variable: switching status $S$}\\ 
		\cmidrule(ll){1-2} 
		Variable & Coefficient (Standard error) \\ 
		\cmidrule(ll){1-2} 
		Intercept & $-8.478 (3.955)^{*}$\\
		BLGFR & $-0.055 (0.030)$ \\
		FPG & $0.974 (0.363)^{**}$\\
		\cmidrule(ll){1-2}  
		Observations & 195 \\
		\bottomrule
	\end{tabular}
	\caption{Logistic regression results for the pooled model for  $P(S=1|\bfL,R=1,\bfC)$, used to estimate the hypothetical estimand.   \\ Predictors: BLGFR, glomerular filtration rate at baseline (ml/min/1.73m$^2$);  FPG, average fasting plasma glucose value  (mmol/l) (for switchers: average before switching, for non-switchers: average before week 26). \\
		Significance codes: $^*p<0.05$, $^{**}p<0.01$ , $^{***}p<0.001$ }
	\label{table: Regression model l tmt group}
\end{table}
	\end{appendices}

\end{document}